\documentclass[AMA,STIX1COL]{WileyNJD-v2}
\usepackage{bm}
\articletype{RESEARCH ARTICLE}%

\begin{document}

\title{Identification of cancer omics commonality and difference via community fusion}

\author[1,2]{Yifan Sun}

\author[3]{Yu Jiang}

\author[1,2,4]{Yang Li}

\author[2,5]{Shuangge Ma}

\authormark{SUN \textsc{et al}}

\address[1]{\orgdiv{Center for Applied Statistics}, \orgname{Renmin University of China}, \orgaddress{\state{Beijing}, \country{China}}}

\address[2]{\orgdiv{School of Statistics}, \orgname{Renmin University of China}, \orgaddress{\state{Beijing}, \country{China}}}

\address[3]{\orgdiv{School of Public Health}, \orgname{University of Memphis}, \orgaddress{\state{Tennessee}, \country{U.S.A}}}

\address[4]{\orgdiv{Statistical Consulting Center}, \orgname{Renmin University of China}, \orgaddress{\state{Beijing}, \country{China}}}

\address[5]{\orgdiv{Department of Biostatistics}, \orgname{Yale University}, \orgaddress{\state{Connecticut}, \country{U.S.A}}}

\corres{Shuangge Ma, 60 College ST, New Haven, CT, 06520.\email{shuangge.ma@yale.edu}\color{black}}

\abstract[Summary]{The analysis of cancer omics data is a ``classic'' problem, however, still remains challenging. Advancing from early studies that are mostly focused on a single type of cancer, some recent studies have analyzed data on multiple ``related'' cancer types/subtypes, examined their commonality and difference, and led to insightful findings. In this article, we consider the analysis of multiple omics datasets, with each dataset on one type/subtype of ``related'' cancers. A Community Fusion (CoFu) approach is developed, which conducts marker selection and model building using a novel penalization technique, informatively accommodates the network community structure of omics measurements, and automatically identifies the commonality and difference of cancer omics markers. Simulation demonstrates its superiority over direct competitors. The analysis of TCGA lung cancer and melanoma data leads to interesting findings.}

\maketitle

\section{Introduction}

For most if not all cancer types, omics profiling studies have been extensively conducted. The early studies are usually focused on a single cancer type/subtype. More recently, as represented by the NCI pan-cancer study, more and more studies have conducted the joint analysis of data on multiple cancer types \cite{weinstein2013cancer,tamborero2013comprehensive}. Such studies can be more challenging and more informative than those on a single cancer type \cite{Guerra2009, Ma2011, Huang2016}. On one hand, with the well-known heterogeneity, differences across cancer types are expected. On the other hand, many studies have suggested the shared omics basis of multiple cancers \cite{Rhodes2004}. As such, a certain level of commonality is also expected. {\it It is important to acknowledge and accommodate both difference and commonality in analysis.}

\begin{figure}
\center
\includegraphics[width=6in]{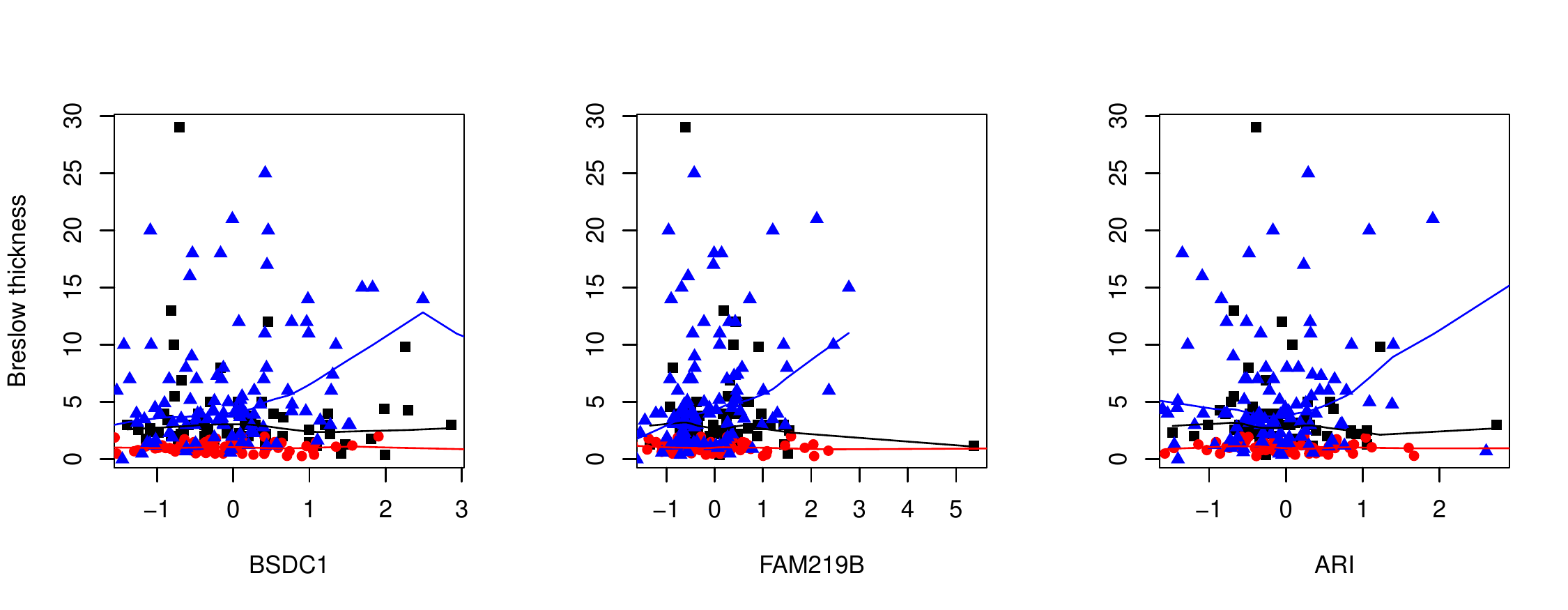}
\caption{Analysis of TCGA melanoma data: Breslow thickness against the expressions of three genes. Three colors correspond to three cancer stages.}\label{fig:gene}
\end{figure}

In the literature, the heterogeneity (difference) across different types of cancers has been sufficiently acknowledged. In contrast, the difference among subjects with the same type of cancer has been noted to a less extent\cite{habermann2007stage, cheng2015stage, palaniappan2016computational}. To fix idea, consider the TCGA melanoma data analyzed in study (for more details, refer to Section \ref{sec:mel}). In Figure \ref{fig:gene}, we show the plots of Breslow thickness (response variable) against the expressions of three genes. The three colors correspond to the three different cancer stages, and the lines are generated using lowess smoothing. It is easy to see that, for different stages of the same cancer, the effects of genes on the response variable are significantly different. It is noted that data on the three stages have been included in one single dataset, and this particular data \cite{akbani2015genomic} as well as those alike \cite{krauthammer2012exome} have usually been analyzed as a whole, with insufficient attention to  heterogeneity/difference. In short, accommodating omics difference associated with a third variable (stage in this particular example) is much needed but insufficiently studied \cite{cheng2015stage}.

Many analytic approaches have been developed for analyzing a single cancer omics dataset\cite{chadeau2013, Ang2016}. 
%The literature is too vast to be reviewed here.
Such methods can be used to analyze, for example, the TCGA melanoma dataset as a whole, which stresses commonality but cannot accommodate difference. They can also be applied to one part of data (which may correspond to one stage, subtype, etc.) at a time, and then results are compared across different data parts to draw conclusions on commonality and difference \cite{palaniappan2016computational}. One major problem is that, since each data part often has a small sample size, results from single-part analysis and hence the final results can be unsatisfactory. Related discussions have been extensively provided in recent integrative analysis studies \cite{Ma2011, Huang2016, Huang2017}. In the literature, the work that is the most relevant to this study is perhaps the contrasted penalization \cite{Shi2014}, which encourages similarity in analysis results (especially regression coefficients) across datasets. Contrasted penalization and some other approaches can be limited by identifying similarity but not commonality, in the sense that parts of the analysis results (from multiple datasets) may be similar but not identical (and thus are not commonly shared).  In addition, they may not sufficiently accommodate the coordination among omics variables.

\begin{figure}
\center
\includegraphics[width=8 cm]{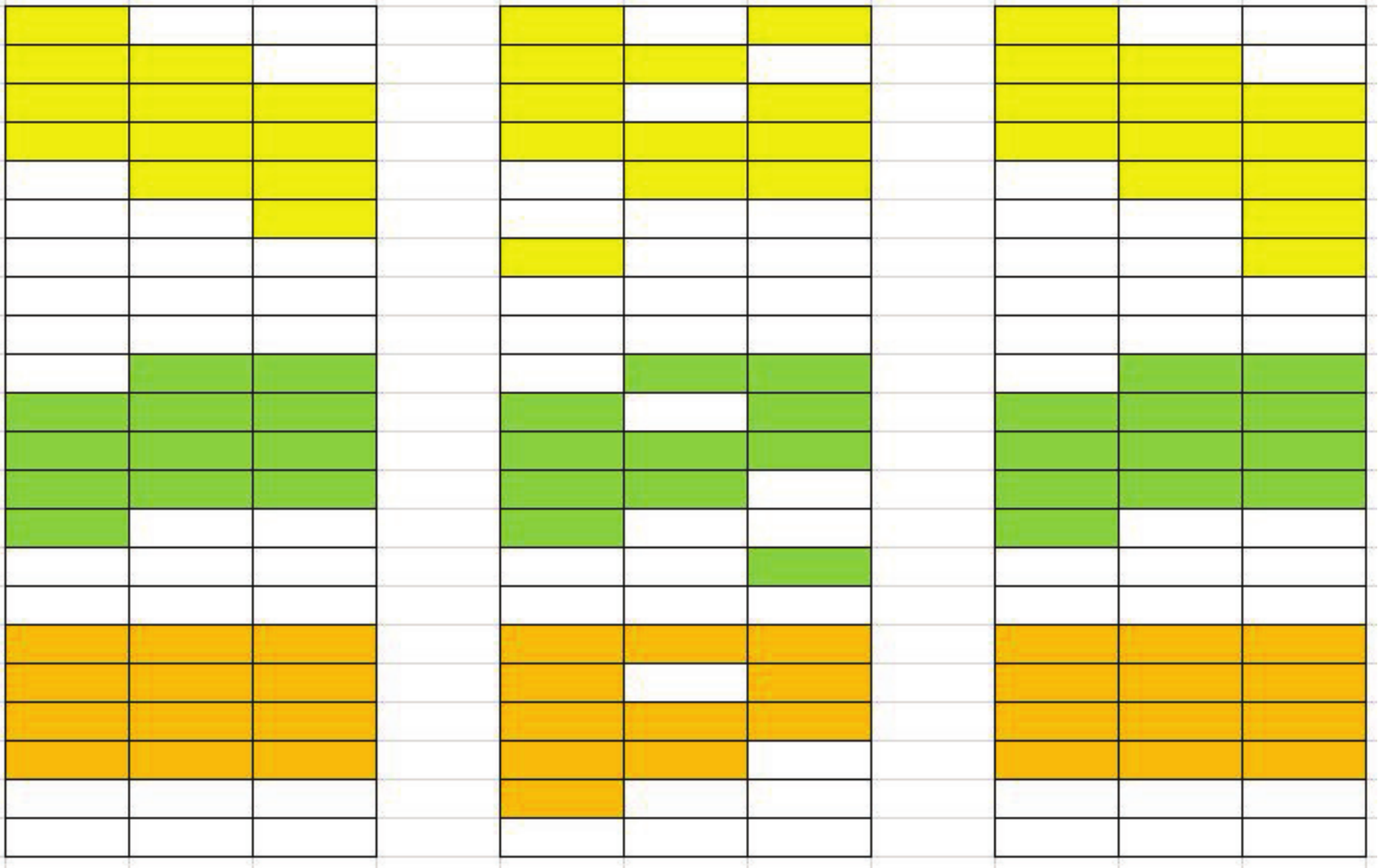}
\caption{Scheme of analysis. Left: true model structure; Middle: analysis by an alternative; Right: CoFu analysis.}\label{graph}
\end{figure}

The analysis scheme of this study is described in Figure \ref{graph}. The left panel describes the true model structure. Here, the three columns correspond to three datasets (with each dataset corresponding to one cancer subtype, stage, etc.), and the rows correspond to genes (or other omics units). The genes form three communities with sizes nine, seven, and six, which are generated from network analysis and describe the coordinated nature of genes. The genes that are associated with the responses are colored. For this schematic example, the orange community behaves the same for the three datasets (commonality), while the yellow and green communities behave differently (difference). {\it Our goal is to conduct community-based analysis (so as to sufficiently accommodate gene coordination) and identify commonality as well as difference in genes' associations with responses}. For such a purpose, a Community Fusion (CoFu) approach is developed. In the middle and right panels of Figure \ref{graph}, the analysis results using an alternative and CoFu are provided, where CoFu has a more accurate identification (more definitive results are provided below in simulation).

For the analysis of cancer omics data, this study has a unique focus on identifying both commonality and difference, and complements the existing studies. A novel CoFu approach is developed. Advancing from methods that pool all datasets together, it can flexibly allow difference across datasets. By jointly analyzing multiple datasets (and hence conducting integrative analysis), it can be more effective than analyzing multiple datasets individually and then pooling results. Advancing from contrasted penalization and other related approaches, it conducts community-based analysis and can identify commonality as opposed to similarity. Overall, this study can deliver a practically useful new venue for analyzing cancer omics data and may lead to important new findings.

\section{Methods}
Consider the analysis of $K$ independent datasets. Each dataset can be on a different type of cancer or a different stage of the same cancer (or be stratified in another meaningful way). In dataset $k(=1, \ldots, K)$, there are $n_k$ i.i.d. observations. Here homogeneity is assumed within but not across datasets. Denote $\bm{y}^{k}=(y_1^k,\ldots,y_{n_k}^k)^\top$ as the response vector and $\bm{X}^k \in \mathbb{R}^{n_k\times p}$ as the design matrix for genes (or other omics units). To simplify notation, it is assumed that the same set of genes is measured in all $K$ datasets. In addition, assume that data processing, for example normalization, has been properly conducted.

For the $k$th dataset, consider the linear regression (LR) model
\begin{equation*}
\label{eq:linear}
\bm{y}^{k}=  \bm{X}^{k}\bm{\beta}^{k}+\bm{\epsilon}^k,\ \ k=1,2,\ldots, K, 
\end{equation*}
where $\bm{\beta}^k=(\beta_1^k,\ldots,\beta_p^k)^\top$ is the length-$p$ vector of regression coefficients, and $\bm{\epsilon}^k$ is the vector of random errors. Here we first consider linear regression, which is the most popular and matches data analyzed in this article. The proposed method is applicable to other models, for example, the GLM model. More details are provided in Appendix \ref{sec:logit}.

For analyzing cancer omics data, network-based analysis, which takes a system perspective and accommodates the interconnections among genes, has been shown to be more informative than individual-gene-based analysis. In our analysis, we accommodate the network community structure, with the understanding that genes in the same community tend to behave in a coordinated manner, while those in different communities tend to behave differently. In recent literature, there have been extensive studies on network and community construction \cite{zhang2017network}. Here we adopt existing construction and assume that the community structure is available prior to analysis. Assume that the $p$ genes belong to $L$ non-overlapping communities, with $p_{(l)}$ genes in community $l$. Denote $\bm{\beta}_{(l)}^k$ as the coefficient vector for community $l$ in dataset $k$.

We propose the CoFu estimate:
\begin{eqnarray}
\label{eq:cri}
&&\{\hat{\bm\beta}^k: k=1,\ldots, K\}=argmin \left\{ \sum_{k=1}^K\frac{1}{2n_k}\parallel \bm{y}^k-\bm{X}^k\bm{\beta}^k\parallel_2^2+
\lambda_1\sum_{k=1}^K\parallel\bm{\beta}^k\parallel_1
\right. \nonumber \\
&& ~~~~~~~~~~~~~~~~~~~~~~~~~~~\left.
+\lambda_2\sum_{k=1}^{K-1}\sum_{l=1}^L\parallel \bm{\beta}^k_{(l)}-\bm{\beta}^{k+1}_{(l)}\parallel_2 \right\},
\end{eqnarray}
where $\lambda_1, \lambda_2>0$ are data-dependent tuning parameters. The nonzero components of $\hat{\bm\beta}^k$ correspond to genes that are associated with the response in dataset $k$. If $\hat{\bm{\beta}}^{k_1}_{(l)}=\hat{\bm{\beta}}^{k_2}_{(l)}$, then genes in community $l$ behave the same in datasets $k_1$ and $k_2$. That is, they represent the commonality shared by the two datasets. Otherwise, they represent the difference. As such, the proposed analysis can simultaneously identify both commonality and difference.

\noindent\underline{Rationale}
In (\ref{eq:cri}), with $K$ independent datasets, the lack-of-fit measure is the sum of $K$ individual ones. Normalization by sample size is taken to avoid domination by larger datasets. The first penalty is Lasso, which has been adopted in a large number of studies. It accommodates the high data dimensionality and conducts regularized estimation and selection of relevant genes. When desirable, it can be replaced by other, more complicated penalties. For example, when it is desirable to accommodate network adjacency (between any two nodes), then Laplacian-type penalties can be further added.

The main advancement is the second penalty, which conducts community-level analysis and has a fusion form. Tailored to data analyzed in Section \ref{sec:realdata}, the penalty has been designed for datasets with a natural order. For example, dataset $k$ may correspond to cancer stage $k$. For two adjacent datasets, the fusion penalty encourages equal regression coefficients, that is, commonality. However, it is flexible and does not reinforce commonality. Different from the ``standard'' fusion penalties, it is imposed on the regression coefficients of different datasets. Different from the contrasted penalties, it conducts community-based analysis: a whole community will be concluded as behaving the same (or differently) in multiple datasets. The results so generated can be more interpretable than those individual-gene-based. In addition, with a nonzero probability, the proposed penalty may generate $\hat{\bm{\beta}}^{k_1}_{(l)}=\hat{\bm{\beta}}^{k_2}_{(l)}$, differing from the contrasted penalization \cite{Shi2014} which generates similar but not identical estimates. For scenarios where a natural order of data does not exist, the second penalty can be revised as $ \lambda_2\sum_{k,j=1,\ldots, K}\sum_{l=1}^L\parallel \bm{\beta}^k_{(l)}-\bm{\beta}^{j}_{(l)}\parallel_2$. Loosely speaking, the newly proposed penalty has a group Lasso form (on the differences of coefficient vectors). It can be potentially replaced by other group penalties (as well as other techniques that can conduct group selection).

\subsection{Computation}
To tackle the non-separability of the penalty, we introduce a new set of parameters: $\bm{\eta}^k=\bm{\beta}^k-\bm{\beta}^{k+1}$. In addition, we also define a set of parameters $\bm{\delta}^k=\bm{\beta}^k$ to separate the $L_1$ norm operation and the sum of squares operation defined on $\bm{\beta}^k$. Then, the minimization in (\ref{eq:cri}) is equivalent to the following constrained optimization problem
\begin{eqnarray*}
\label{eq:optimization}
\min f(\bm{\beta},\bm{\eta},\bm{\delta})   \equiv \sum_{k=1}^K\frac{1}{2n_k}\parallel \bm{y}^k-\bm{X}^k\bm{\beta}^k \parallel_2^2+\lambda_1\sum_{k=1}^K\parallel\bm{\delta}^k
\parallel_1+\lambda_2\sum_{k=1}^{K-1}\sum_{l=1}^L\parallel \bm{\eta}_{(l)}^k \parallel_2,
\end{eqnarray*}
subject to
\[\bm{\beta}^k-\bm{\beta}^{k+1}-\bm{\eta}^k=0, \ \ k=1,2,\ldots, K-1,  \]
and
\[\bm{\beta}^k=\bm{\delta}^k, \ \ k=1,2, \ldots, K.\]
Here $\bm{\beta}=((\bm{\beta}^1)^\top, \ldots,(\bm{\beta}^K)^\top)^\top$,  $\bm{\eta}=((\bm{\eta}^1)^\top,\ldots,(\bm{\eta}^{K-1})^\top)^\top$, and $\bm\delta=((\bm\delta^1)^\top,\ldots,(\bm\delta^K)^\top)^\top$.
By the augmented Lagrangian method\cite{Hestenes1969}, the estimates can be obtained by minimizing
\begin{eqnarray*}
\label{Lagrange}
\mathcal{L}(\bm{\beta},\bm{\eta},\bm{\delta}, \bm{u},\bm{v}) &= & f(\bm{\beta},\bm{\eta},\bm{\delta}) +\sum_{k=1}^{K-1} (\bm{u}^k)^\top(\bm{\beta}^k-\bm{\beta}^{k+1}-\bm{\eta}^k)+
\sum_{k=1}^K(\bm{v}^k)^\top (\bm{\beta}^k-\bm{\delta}^k)\nonumber \\
&& +\frac{\sigma}{2}\Big(\sum_{k=1}^{K-1}\parallel \bm{\beta}^k-\bm{\beta}^{k+1}-\bm{\eta}^k \parallel_2^2+\sum_{k=1}^K\parallel \bm{\beta}^k-\bm{\delta}^k \parallel_2^2\Big),
\end{eqnarray*}
where the dual variables $\bm{u}=((\bm{u}^1)^\top,\ldots,(\bm{u}^{K-1})^\top)^\top$ and $\bm{v}=((\bm{v}^1)^\top,\ldots,(\bm{v}^K)^\top)^\top$ are Lagrange multipliers, and $\sigma>0$ is the penalty parameter. Throughout this article, we set $\sigma=2$, which leads to satisfactory numerical results.

We compute the estimate of $(\bm\beta,\bm{\eta},\bm{\delta}, \bm{u},\bm{v})$, denoted as $(\hat{\bm\beta},\hat{\bm\eta},\hat{\bm\delta},\hat{\bm u},\hat{\bm v})$, iteratively by using the ADMM method. For a given $(\bm{\eta},\bm{\delta}, \bm{u},\bm{v})$, to obtain the update of $\bm{\beta}$, we set 
$\partial L/\partial \bm\beta$ to be zero, where
\begin{eqnarray*}
L&=&\sum_{k=1}^K\frac{1}{2n_k}\parallel\bm{X}^k \bm{\beta}^k-\bm{y}^k \parallel_2^2\nonumber \\
& & +\frac{\sigma}{2}\Big(\sum_{k=1}^{K-1}\parallel \bm\beta^k-\bm\beta^{k+1}-\bm\eta^k+\bm u^k/\sigma\parallel_2^2+\sum_{k=1}^K\parallel \bm\beta^k-\bm\delta^k +\bm v^k/\sigma\parallel_2^2\Big)\nonumber\\
&=& \frac{1}{2}\parallel\bm{X} \beta-\mathbf{y} \parallel_2^2+\frac{\sigma}{2}\Big(\parallel \bm{A}\bm\beta-\bm\eta+\bm u/\sigma\parallel_2^2+\parallel\bm\beta-\bm\delta+\bm v/\sigma\parallel_2^2\Big).
\end{eqnarray*}
$\bm{X}={\rm{diag}}(\bm{X}^1/\sqrt{n_1},\dots,\bm{X}^k/\sqrt{n_k})$, $\bm{y}=((\bm{y}^1/\sqrt{n_1})^\top,\ldots,
(\bm{y}^K/\sqrt{n_k})^\top)^\top$. $e_j$ is the length-$K$ column vector, with the $j$th element equal to 1 and the rest equal to 0. $\Delta=\{e_j-e_{j+1}, j=1,\ldots,K-1\}^\top$. $\bm{A}=\Delta \otimes \bm{I}_p$, where $\bm{I}_p$ denotes the $p\times p$ identity matrix, and $\otimes$ denotes the Kronecker product.

Thus, for a given $(\bm\eta(t),\bm \delta(t), \bm u(t),\bm v(t))$ at the $t$th iteration, 
\begin{eqnarray}
\label{up:beta}
\bm\beta(t+1)  =    [\bm X^\top \bm X+\sigma(\bm{A}^\top \bm A+\bm I_{pK})]^{-1}
 \times \{\bm X^\top \bm y+\sigma[\bm A^\top(\bm\eta(t)-\bm u(t)/\sigma)+\bm \delta(t)-\bm v(t)/\sigma]\},
\end{eqnarray}
where $\bm{I}_{pK}$ denotes the $pK\times pK$ identity matrix.

The component of the Lagrange function $\mathcal{L}(\bm{\beta},\bm{\eta},\bm{\delta}, \bm{u},\bm{v})$ that depends on $\bm\delta$ is
\begin{equation*}
\lambda_1 \parallel \bm\delta \parallel_1+\frac{\sigma}{2}\parallel \bm\beta-\bm\delta+\bm v/\sigma \parallel_2^2.
\end{equation*}
Hence, the closed-form solution for $\bm\delta$ is
\begin{equation*}
\hat{\bm\delta}=\mathrm{ST}_{\lambda_1/\sigma}(\bm\beta+\bm v/\sigma),
\end{equation*}
where $\mathrm{ST}_{\lambda}(x)=\mathrm{sign}(x)(|x|-\lambda)_+$ is the soft thresholding function, and $(x)_+=x$ if $x>0$, and $(x)_+=0$ otherwise. Then, the update of $\bm \delta$ at the $(t+1)$th iteration is
\begin{equation}
\label{up:delta}
\bm\delta(t+1)=\mathrm{ST}_{\lambda_1/\sigma}[\bm\beta(t+1)+\bm v(t)/\sigma].
\end{equation}
For $\bm\eta^k$, the relevant component in $\mathcal{L}(\bm{\beta},\bm{\eta},\bm{\delta},\bm{u},\bm{v})$ is
\begin{equation}
\label{eq:eta}
\lambda_2\parallel \bm\eta^k_{(l)}\parallel_2+\frac{\sigma}{2} \parallel \bm\beta^{k}_{(l)}- \bm\beta^{k+1}_{(l)}-\bm\eta^k_{(l)}+\bm u^{k}_{(l)}/\sigma\parallel_2^2.
\end{equation}
The minimizer of (\ref{eq:eta}) is
\begin{equation*}
\hat{\bm\eta}^k_{(l)}=\Big( 1-\frac{\lambda_2}{\sigma\parallel \bm\beta^k_{(l)}-\bm\beta^{k+1}_{(l)}+\bm u^k_{(l)}/\sigma\parallel_2} \Big)_{+} (\bm\beta^k_{(l)}-\bm\beta^{k+1}_{(l)}+\bm u^k_{(l)}/\sigma), \ k=1,2,\ldots, K-1.
\end{equation*}
We can thus obtain the update of $\bm \eta^k_{(l)}$ at the $(t+1)$th iteration as 
\begin{eqnarray}
\label{up:eta}
\bm \eta^k_{(l)}(t+1)  =  \Big( 1-\frac{\lambda_2}{\sigma\parallel \bm\beta^k_{(l)}(t+1)-\bm\beta^{k+1}_{(l)}(t+1)+\bm u^k_{(l)}(t)/\sigma\parallel_2} \Big)_{+}
\times [\bm \beta^k_{(l)}(t+1)-\bm \beta^{k+1}_{(l)}(t+1)+\bm u^{k}_{(l)}(t)/\sigma].
\end{eqnarray}
Finally, the estimates of $\bm u$, $\bm v$ are updated as
\begin{eqnarray}
&& \bm u(t+1) = \bm u(t)+\sigma[\bm A \bm \beta(t+1)-\bm \eta(t+1)],\label{up:u}\\
&&\bm v(t+1) =  \bm v(t)+\sigma[\bm \beta(t+1)-\bm \delta(t+1)].\label{up:v}
\end{eqnarray}

\noindent\underline{Overall algorithm}
The overall algorithm proceeds as follows:

\noindent Step 1: Initialize $\bm \beta(0)$. Let $\bm \delta(0)=\bm \beta(0)$, $\bm \eta(0)=\bm A\bm \beta(0)$, $\bm u(0)=0$, and $\bm v(0)=0$. In our numerical study, we use the Lasso estimate (without fusion) as the initial value.

\noindent Step 2: At the $(t+1)$th iteration, compute $\bm \beta(t+1), \bm \delta(t+1), \bm \eta(t+1), \bm u(t+1)$ and $\bm v(t+1) $ according to Eqs. (\ref{up:beta}), (\ref{up:delta}), (\ref{up:eta}), (\ref{up:u}), and (\ref{up:v}), respectively.

\noindent Step 3: Repeat Step 2 until convergence. Specifically, convergence is concluded if all of the primal residuals and dual residuals are small enough, that is,
\begin{equation*}
\max\{\parallel \bm{A}\bm{\beta(t+1)}-\bm{\eta(t+1)} \parallel_2 , \parallel \bm{\beta(t+1)}-\bm{\delta(t+1)} \parallel_2,\parallel \bm{\delta(t+1)}-\bm{\delta(t)} \parallel_2, \parallel \bm{\eta(t+1)}-\bm{\eta(t)} \parallel_2\}<\epsilon.
\end{equation*}
In numerical study, $\epsilon$ is set to be $10^{-3}$.

\noindent\underline{Tuning parameter selection}
The proposed approach involves two tuning parameters: $\lambda_1$ and $\lambda_2$. In numerical study, they are selected using V-fold cross validation (CV). In this article, $V=5$. More specifically, each dataset is partitioned randomly into 5 non-overlapping subsets with equal sizes. We apply a two-dimensional grid search for $\lambda_1$ and $\lambda_2$ with  $\lambda_2\in(0.001,0.01,0.1,1)$. Let $\lambda_1^{\text{max}}$ be the minimal $\lambda$ such that all regression coefficients shrink to 0, i.e., $\lambda_1^{\max}=\max_{k} \parallel \frac{\bm{(X^k)}^\top\bm{y}^k}{n^k} \parallel_{\infty}$. We choose $\lambda_1^{\min}$ to be some small fraction (default value is 0.01 in our implementation) of $\lambda_1^{\max}$ and log-linearly interpolate between $\lambda_1^{\min}$ and $\lambda_1^{\max}$. For the CoFu method, the CV is computationally affordable. For example, the 5-fold CV for a simulated dataset takes less than 15 minutes on a desktop PC.

\section{Simulation}
\label{sec:sim}
In simulation, we set $K=3$, $n_k=200$, and $p=1,000$. Genes form $L=50$ non-overlapping communities, with community sizes ranging from $\frac{3p}{4L}$ to $\frac{5p}{4L}$. As in the literature \cite{Huang2017}, we simulate omics measurements from multivariate normal distributions, which may mimic gene expression data (as analyzed below). The normal distributions have marginal means 0, marginal variances 1, and correlation matrix $\Sigma$. The following correlation structures are considered: (a) structured correlation. Omics measurements within the same communities are more strongly correlated than those in different communities. More details are provided below; (b) unstructured correlation. The correlation coefficient between measurements $i$ and $j$ is randomly sampled from the uniform distribution $\mathcal{U}[0.2,1]$; and (c) no correlation. That is, all omics measurements are independent. This may serve as a test of sensitivity and examine performance of the proposed approach when there is a lack of well defined network structure.

Roughly speaking, the structured correlation matrix corresponds to the scenario where two genes within the same community are strongly correlated while two genes within different communities are weakly correlated. Since in reality it is possible that not all genes within the same community are correlated, we generate the structured correlation matrix via network-based analysis, which is considerably more complicated than in the literature and consists of the following steps: (a) generate an unweighted network with a community structure\footnote{Most networks of interest display community structures, that is, their nodes are organized into groups, called communities. The density of edges within a community is comparatively higher than that between communities.}, (b) add a weight to each edge, which quantifies the connection between two nodes (omics measurements), and (c) generate the correlation matrix $\Sigma$. More specifically, we adopt the degree-correlated stochastic block model (DC-SBM) \cite{Karrer2011} to generate an unweighted network with a community structure. To mimic networks in reality, we generate a degree sequence of nodes following the power-law distribution with exponent $\gamma$. For each pair of nodes $i$ and $j$, an undirected edge is placed with probability
 \begin{equation}
\label{pij_unnorm}
p_{ij}=\frac{<d>p}{Z}d_i d_j q_{m_im_j},
\end{equation}
where $Z=\sum_{i,j}d_i d_j q_{m_im_j}$ is the normalization constant, $d_i$ is the degree of node $i$, $<d>$ is the average degree of nodes, and $m_i$ represents the community that node $i$ belongs to. $q_{m_im_j}$ represents the connection probability between communities  $m_i$ and $m_j$. If $m_1=m_2$, $q_{m_1m_2}$ is sampled from an uniform distribution $\mathcal{U}[0.3,0.5]$, otherwise $q_{m_1m_2}=0.02$. In all simulations, we set $\gamma=2.5$ and $<d>=10$.
Next, we add a weight to each edge. Specifically, for edges between nodes in the same community and edges between nodes in different communities, weights are sampled independently from uniform distributions $\mathcal{U}[0.5, 1]$ and $\mathcal{U}[0.2, 0.5]$, respectively. Denote the adjacency matrix of the weighted network as $\Sigma_0$, with all diagonal elements equal to 1. Note that $\Sigma_0$ is not necessarily positive definite. To guarantee positive definiteness, we set  $\Sigma=\Sigma_0-(\lambda_{\min}-\frac{1}{p})\bm{I}_{p}$, where $\lambda_{\min}$ is the smallest eigenvalue of $\Sigma_0$, and $\bm{I}_p$ is the $p\times p$ identity matrix. For a more intuitive presentation, we plot one simulated structured correlation matrix in Figure \ref{fig:sigma} (Appendix \ref{sec:table}) with $p=200$, $L=10$, and other parameters as described above. As shown in the figure, only genes that are connected by an edge are correlated. The correlation between two connected genes within the same community is strong, whereas that between two connected genes from different communities is weak. 

Each dataset has $r$ important omics measurements with $r=100$ and 150, which are distributed uniformly across communities. All communities are divided randomly into three categories: (a) all-overlapping, where the three datasets have the same sparsity structure and also the same regression coefficients; (b) half-overlapping, where datasets 1 and 2 share half of the important effects, for which the regression coefficients are identical. The same is true for datasets 2 and 3. Half of the important effects in datasets 2 and 3 are dataset-specific; and (c) non-overlapping, where there is no important effect shared by any two datasets. Denote $\rho_a$, $\rho_h$, $\rho_n$ ($\rho_a+\rho_h+\rho_n=1$) as the proportions of all-, half-, and non-overlapping communities, respectively. For the important effects, their regression coefficients are (a) all set to be 0.5; (b) sampled from the uniform distribution $\mathcal{U}[0.2,1]$; and (c) sampled from different distributions. Specifically, the nonzero regression coefficients in dataset 2 are from $\mathcal{U}[0.4,0.7]$, those specific to dataset 1 are from $\mathcal{U}[0.1,0.3]$, and those specific to dataset 3 are from $\mathcal{U}[0.8,1]$. The random errors are simulated independently from $N(0,1)$. The response variables are computed from the linear models.

\begin{table}[htbp]
   \centering
   \caption{
   Simulation under the LR model: mean(sd) of AUC for effect identification. All nonzero coefficients are equal to 0.5.} % requires the topcapt package
   \begin{tabular}{lcccccc} % Column formatting, @{} suppresses leading/trailing space
      \toprule
      %\multicolumn{2}{c}{correlation} \\
      & & $r=100$ & &  & $r=150$\\
      \cmidrule(r){2-4} \cmidrule(l){5-7}% Partial rule. (r) trims the line a little bit on the right; (l) & (lr) also possible
         & structured &unstructured & independence  &structured &unstructured & independence \\
      \midrule
      \multicolumn{3}{l}{$(\rho_a,\rho_h,\rho_n)=(0.1,0,0.9)$} \\
        P.Lasso & 0.619(0.014)&0.593(0.015)&0.62(0.013)&0.578(0.017)&0.574(0.013)&0.579(0.018)\\
       S.Lasso  & 0.77(0.009)&0.833(0.01)&0.761(0.012)&0.695(0.015)&0.777(0.01)&0.685(0.012)\\
       CoFu &0.771(0.01)&0.813(0.012)&0.753(0.012)&0.694(0.012)&0.762(0.012)&0.676(0.013)\\
       \multicolumn{3}{l}{$(\rho_a,\rho_h,\rho_n)=(0.1,0.9,0)$} \\
        P.Lasso &0.732(0.015)&0.743(0.02)&0.725(0.015)&0.621(0.02)&0.703(0.012)&0.646(0.021)\\
       S.Lasso  &0.781(0.013)&0.85(0.012)&0.782(0.012)&0.689(0.013)&0.787(0.007)&0.687(0.013)\\
       CoFu &0.827(0.013)&0.891(0.008)&0.814(0.009)&0.73(0.012)&0.8(0.012)&0.716(0.015)\\
      \multicolumn{3}{l}{$(\rho_a,\rho_h,\rho_n)=(0.2,0.6,0.2)$} \\
       P.Lasso & 0.752(0.017) &0.719(0.023) &0.744(0.017) &0.689(0.018) &0.666(0.014) &0.68(0.016)\\
       S.Lasso  & 0.772(0.011) & 0.847(0.011) & 0.76(0.011) & 0.699(0.012) &0.786(0.008) &0.689(0.01)\\
       CoFu  &0.838(0.012) &0.878(0.007) &0.826(0.01) & 0.762(0.008) &0.803(0.01) &0.753(0.011)\\
       \multicolumn{3}{l}{$(\rho_a,\rho_h,\rho_n)=(0.4,0.1,0.5)$} \\
       P.Lasso & 0.746(0.012) &0.741(0.012) &0.739(0.012) &0.698(0.013) &0.684(0.01) &0.69(0.014)\\
       S.Lasso  & 0.775(0.007) & 0.831(0.013) & 0.752(0.013) &0.691(0.016) &0.775(0.008) &0.681(0.017)\\
       CoFu &0.823(0.009) &0.865(0.011) &0.817(0.008) &0.756(0.017) &0.795(0.011) &0.746(0.016)\\
       \multicolumn{3}{l}{$(\rho_a,\rho_h,\rho_n)=(0.5,0.5,0)$} \\
      P.Lasso & 0.885(0.013) &0.866(0.007) &0.885(0.01) &0.82(0.019) &0.799(0.017) &0.812(0.017)\\
       S.Lasso  & 0.776(0.011) &0.834(0.009)& 0.764(0.015) & 0.692(0.015) &0.781(0.009) &0.69(0.009)\\
       CoFu &0.894(0.011) &0.893(0.005) &0.891(0.011) &0.817(0.009) &0.839(0.011)&0.816(0.01) \\
       \multicolumn{3}{l}{$(\rho_a,\rho_h,\rho_n)=(0.6,0.2,0.2)$} \\
        P.Lasso & 0.878(0.011) & 0.847(0.013) &0.873(0.009) & 0.831(0.022) &0.779(0.016) &0.809(0.024) \\
       S.Lasso  &0.762(0.012) &0.831(0.011) &0.76(0.013) & 0.692(0.016) & 0.777(0.009) & 0.684(0.018) \\
       CoFu &0.887(0.011) &0.888(0.009) &0.881(0.008) &0.822(0.017) &0.829(0.011) &0.827(0.019)\\
        \multicolumn{3}{l}{$(\rho_a,\rho_h,\rho_n)=(0.9,0,0.1)$} \\
        P.Lasso &0.89(0.014)&0.925(0.02)&0.872(0.015)&0.832(0.017)&0.884(0.012)&0.828(0.02)\\
       S.Lasso  &0.768(0.009)&0.845(0.012)&0.774(0.012)&0.691(0.013)&0.785(0.008)&0.685(0.015)\\
       CoFu &0.876(0.013)&0.899(0.008)&0.859(0.01)&0.805(0.012)&0.864(0.012)&0.798(0.014)\\
      \bottomrule
   \end{tabular}
   %\caption{Remember, \emph{never} use vertical lines in tables.}
   \label{tab:var1}
\end{table}

\begin{table}[htbp]%\small
   \centering
   \caption{
   Simulation under the LR model: mean(sd) of AUC for community differentiation. All nonzero coefficients are equal to 0.5.
} % requires the topcapt package
   \begin{tabular}{lcccccc} % Column formatting, @{} suppresses leading/trailing space
      \toprule
      %\multicolumn{2}{c}{correlation} \\
      & & $r=100$ & &  & $r=150$\\
      \cmidrule(r){2-4} \cmidrule(l){5-7}% Partial rule. (r) trims the line a little bit on the right; (l) & (lr) also possible
         & structured &unstructured & independence  &structured &unstructured & independence \\
      \midrule
      \multicolumn{3}{l}{$(\rho_a,\rho_h,\rho_n)=(0.1,0,0.9)$} \\
       S.Lasso  &0.578(0.035)&0.54(0.097)&0.571(0.066)&0.565(0.078)&0.515(0.105)&0.537(0.068)\\
       CoFu &0.739(0.033)&0.762(0.038)&0.724(0.042)&0.711(0.042)&0.74(0.042)&0.679(0.039)\\
       \multicolumn{3}{l}{$(\rho_a,\rho_h,\rho_n)=(0.1,0.9,0)$} \\
       S.Lasso  &0.543(0.057)&0.54(0.092)&0.53(0.066)&0.496(0.088)&0.468(0.105)&0.484(0.071)\\
       CoFu &0.712(0.038)&0.762(0.036)&0.724(0.043)&0.692(0.042)&0.691(0.062)&0.656(0.052)\\
      \multicolumn{3}{l}{$(\rho_a,\rho_h,\rho_n)=(0.2,0.6,0.2)$} \\
       S.Lasso  & 0.573(0.082) & 0.54(0.105) & 0.549(0.074) & 0.517(0.066) &0.52(0.119) &0.509(0.081)\\
       CoFu &0.744(0.032) &0.721(0.05) &0.704(0.048) & 0.679(0.049) &0.71(0.053) &0.679(0.042)\\
       \multicolumn{3}{l}{$(\rho_a,\rho_h,\rho_n)=(0.4,0.1,0.5)$} \\
       S.Lasso  & 0.565(0.043) & 0.541(0.091) & 0.557(0.061) &0.538(0.063) &0.531(0.1) &0.552(0.06)\\
       CoFu &0.749(0.031) &0.746(0.032) &0.752(0.038) &0.724(0.049) &0.737(0.066) &0.73(0.057)\\
       \multicolumn{3}{l}{$(\rho_a,\rho_h,\rho_n)=(0.5,0.5,0)$} \\
       S.Lasso  & 0.533(0.048) & 0.49(0.103) & 0.552(0.062) & 0.476(0.099) &0.492(0.124) &0.513(0.084)\\
       CoFu &0.777(0.043) &0.771(0.029) &0.779(0.046) &0.728(0.038) &0.724(0.036)&0.729(0.03) \\
       \multicolumn{3}{l}{$(\rho_a,\rho_h,\rho_n)=(0.6,0.2,0.2)$} \\
       S.Lasso  & 0.515(0.036) & 0.5(0.096) & 0.53(0.043) &0.438(0.093) &0.477(0.1) &0.486(0.087)\\
       CoFu &0.755(0.053) &0.755(0.041) &0.779(0.04) &0.685(0.051) &0.721(0.046) &0.694(0.033)\\
        \multicolumn{3}{l}{$(\rho_a,\rho_h,\rho_n)=(0.9,0,0.1)$} \\
       S.Lasso  &0.518(0.066)&0.498(0.098)&0.54(0.07)&0.466(0.078)&0.441(0.098)&0.45(0.076)\\
       CoFu &0.965(0.042)&0.965(0.048)&0.975(0.042)&0.947(0.045)&0.907(0.048)&0.961(0.048)\\
      \bottomrule
   \end{tabular}
   %\caption{Remember, \emph{never} use vertical lines in tables.}
   \label{tab:group1}
\end{table}

We compare CoFu with 
%To better gauge performance of the proposed approach, 
two closely related alternatives: (a) P.Lasso, which pools all datasets together and applies Lasso. This approach emphasizes commonality but cannot detect difference across datasets; and (b) S.Lasso, which applies Lasso to each dataset separately, and then the results are combined and compared. This is virtually a meta-analysis strategy, allows difference, but cannot encourage commonality. 
For CoFu and the alternatives, we are mainly interested in two identification accuracy. The first, as in many other studies, is the identification of nonzero effects. The second, which is unique to this study, is the identification of communities that behave the same/differently in different datasets. More precisely, for a given community, if the $L_2$ norm of the difference between the regression coefficient vectors of two adjacent datasets is less than 0.01, we conclude commonality; otherwise, difference is concluded. The identified commonality/difference is compared against the true. For the proposed as well as alternatives, tuning parameter values affect identification performance. To get a more comprehensive view and minimize the (possibly different) impact of tuning on different methods, we follow the literature, consider a sequence of tunings, and evaluate using the ROC (receiver operating characteristic) approach, under which the AUC is the measure of identification accuracy (more details in Appendix \ref{sec:AUC}). Considering that in practice a definitive set of results may be desirable, we also evaluate identification results using true/false positive rates with tunings selected using 5-fold CV. In addition, we evaluate estimation and prediction performance. Specifically, estimation is quantified by using ERMSE (estimation RMSE), which is defined as $\sqrt{\sum_k \parallel \bm{\beta}^k-\hat{\bm{\beta}}^k\parallel^2_2}$, and prediction is quantified by using  PRMSE (prediction RMSE), which is defined as $\sqrt{\sum_k \parallel \bm{y}^k-\bm{X}^k\hat{\bm{\beta}}^k\parallel^2_2}$.

We simulate 100 replicates for each setting. For the settings with all nonzero coefficients equal to 0.5, we show the identification of nonzero effects in Table \ref{tab:var1} and differentiation of communities with commonality/difference in Table \ref{tab:group1}, where the AUC summaries are presented. Results for the other settings are shown in Appendix \ref{sec:table}. The proposed CoFu is observed to have favorable performance in identifying nonzero effects. Consider for example Table \ref{tab:var1}. With $r=100$, $(\rho_a,\rho_h,\rho_n)=(0.4,0.1,0.5)$, and the structured correlation, the mean AUCs are 0.746 (P.Lasso), 0.775 (S.Lasso), and 0.823 (CoFu), respectively. Performance of S.Lasso is not strongly affected by the overlapping across datasets, as it analyzes each dataset separately. In general, S.Lasso has good performance, however, is inferior to CoFu under most settings. P.Lasso has superior performance when the sets of important effects in the three datasets almost entirely overlap, for example, when $(\rho_a,\rho_h,\rho_n)=(0.9,0,0.1)$. However, as expected, its performance deteriorates significantly when there are large differences across datasets. It is noted that CoFu still has competitive performance even under the worst case scenario, thus providing a ``safe'' choice in practice. Similar observations are also made under the other settings as presented in Appendix \ref{sec:table}. In the differentiation of community commonality/difference, note that as P.Lasso is not capable of identifying differences across datasets, its results are not presented. Simulation shows that CoFu significantly outperforms S.Lasso. For example in Table \ref{tab:group1} with $r=100$ and $(\rho_a,\rho_h,\rho_n)=(0.1,0,0.9)$, CoFu has AUCs 0.739 (structured), 0.762 (unstructured), and 0.724 (independence), respectively, while S.Lasso has AUCs below 0.6. For the settings presented in Appendix \ref{sec:table}, the observed patterns are similar. The results with tunings selected using CV are presented in Table \ref{tab:CV1}-\ref{tab:CV3} (Appendix \ref{sec:table}). The CoFu method is observed to have superior performance in terms of identification, estimation, and prediction. 

Simulation is also conducted under the Logit model, a representative of generalized linear models. Details of the estimation procedure are presented in Appendix \ref{sec:logit}. In simulation, covariates are generated in the same way as described above. The response values are generated from the Logit model and Bernoulli distribution. The identification results measured by AUCs are presented in Table \ref{tab:log1}-\ref{tab:log2} (Appendix \ref{sec:table}). The CoFu method is observed to have similar superior identification performance as under the LR model. It is noted that the improvement over the alternatives may not be as large as under the LR model. To this end, we conduct a nonparametric test on the paired AUC values and find that the improvement is statistically significant for all scenarios. For example, for $r=100$, $(\rho_a,\rho_h,\rho_n)=(0.1,0,0.9)$, and from $\mathcal{U}[0.2,1]$, the differences between the CoFu's and S.Lasso's AUC values have p-values 0.002 and $<10^{-9}$ for nonzero effect and community identification, respectively. 

\section{Data analysis}
\label{sec:realdata}
TCGA (The Cancer Genome Atlas) is a collaborative effort organized by NCI and has recently published high-quality profiling data on multiple cancer types. The analysis of TCGA data has led to interesting findings. In our analysis, both clinical and genetic data are downloaded from the cBioPortal website.

\subsection{Analysis of cutaneous melanoma data}
\label{sec:mel}
We first consider the SKCM (cutaneous melanoma) data \cite{akbani2015genomic}. As in the literature \cite{jiang2016integrated}, the inclusion criteria are: (1) white patients, (2) no neo-adjuvant therapy before tumor sample collection, (3) the type of skin upon which melanoma arose is non-glabrous skin, (4) no missing values in Breslow thickness and AJCC pathologic tumor stage, and (5) with gene expression measurements. In our analysis, we are interested in the regulation of Breslow thickness, which is an important prognostic marker, by gene expressions. In published studies \cite{chai2017analysis}, similar analysis has been conducted, however, {\it using samples of all tumor stages and with insufficient attention to the potential difference across stages}. Partly motivated by Figure \ref{fig:gene}, our ``hypothesis'' is that the regulation relationships for different stages have commonality as well as difference. There are a total of 240 samples, with 70 in stage I, 60 in stage II, and 110 in stages III and IV.

A total of 18,947 gene expression measurements are available. From the KEGG pathway database downloaded from the Broad Institute, we identify 5,266 unique genes, representing 186 pathways. Matching those gene names with those in the SKCM dataset, we identify 4,243 genes for downstream analysis. Although in principle it is possible to directly apply the proposed approach to these genes, to obtain more reliable analysis results, we further conduct a supervised screening. Specifically, we compute the Pearson correlation coefficient of each gene with the response variable and identify those with p-values less than 0.05. A total of 973 genes are identified. We construct a dense network based on correlations and then generate a sparse one by filtering out edges that are not statistically significant at the 0.05 level \cite{Serrano2009}. The resulted network has 15,891 edges (detailed structure available from the authors). The Louvain method \cite{Blondel2008}, which performs a greedy optimization of community identification in a hierarchical manner, is applied and identifies 46 communities.

\begin{figure}[h] %
   \centering
   \includegraphics[width=7.5in]{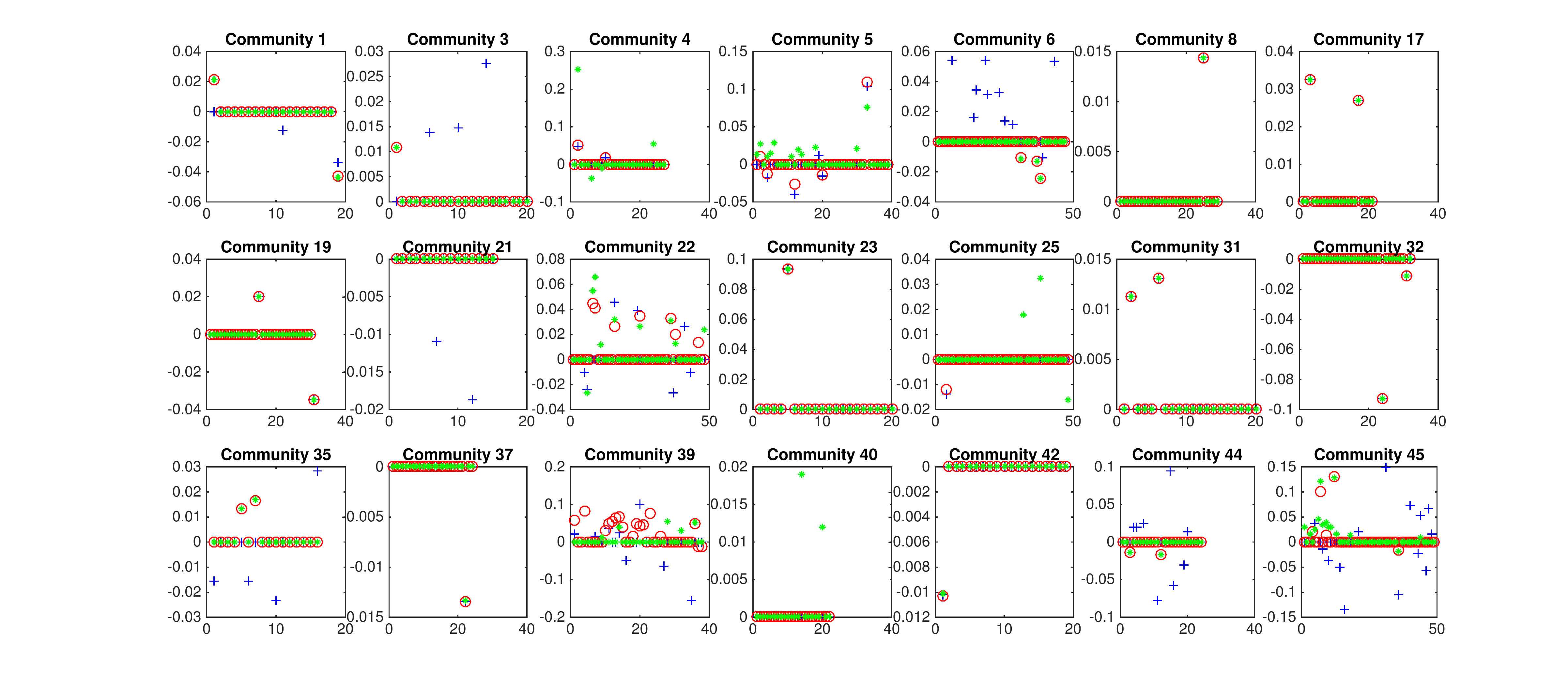}
   \caption{Analysis of the TCGA SKCM data. Blue crosses correspond to stage I, red circles to stage II, and green filled circles to stage III and IV.}
   \label{fig:Mela}
\end{figure}

The analysis results of CoFu are summarized in Figure \ref{fig:Mela}. Briefly, a total of 21 communities, with 126 genes, are identified as associated with the response. Among them, 8 communities behave the same across the three stages, and the rest behave differently. More detailed estimation results are available from the authors. Simply eyeballing Figure \ref{fig:Mela} suggests that some communities, for example 22, 39, and 45, demonstrate significantly different stage-specific properties. Such differences have not been well noted in the literature and deserve additional attention. Literature search suggests that the CoFu identified genes may have important implications. For example, the gene that has the strongest signal in all three stages is ARMC, armadillo repeat containing 2. It belongs to a family of Armadillo repeat proteins, which play important roles in cell-cell adhension, cytoskeletal regulation and intracellular signaling. Other genes that also have strong signals in all three stages include C10orf114 and KTGNR. C10orf114 is also known as CASC10. Over expression of C10orf114 is associated with poor survival in glioma and urothelial cancer (www.proteinatlas.org/ENSG00000204682-CASC10/cell). KTGNR (DNAH5) encodes an axonemal heavy chain of dynein proteins. It works as a force-generating protein with ATPase activity. The TRA2B-DNAH5 fusion has been identified as a novel oncogenic driver in lung cancer \cite{Li2016}. In addition to the genes that are associated with Breslow depth in all three stages, we have also identified genes that have strong signals only in the earlier or later stages. The top three genes that only have strong signals in stage I are DKFZP434B094, AK3 and ANKS1A. AMN and PDCR are found to have strong signals only in stage II. FAM219B and BSDC1 are found to be stage III and IV specific. In addition, genes that are more associated with Breslow depth in later stages are LOC392331, ANKRD20A20P, CRELD2, CEBPG, and others. CEBPG is one of the C/EBP transcription factors that regulate cell growth and differentiation of various tissues. One study has shown that CEBPG is a suppressor of myeloid differentiation in acute myeloid leukemia \cite{Alberichjord2012}.

Different findings are generated by the alternatives. The estimated coefficients are plotted in Figures \ref{fig:lasso_Mela} (S.Lasso) and \ref{fig:single_Mela} (P.Lasso) in Appendix \ref{sec:table}. S.Lasso identifies 38 communities, with 176 genes, as associated with the response variable, and P.Lasso identifies 39 communities, with 105 genes. With their particular properties, S.Lasso identifies all communities (with nonzero effects) as behaving differently across stages, while P.Lasso identifies all communities as behaving the same. Biologically speaking, the CoFu results, which have both commonality and difference, are more sensible.

We also evaluate prediction performance and stability of each method. Specifically, each dataset is randomly divided into a training and testing set, with sizes 2:1. The regression parameters are estimated only using the training set and used to make prediction for the testing set. We use the root mean square error (RMSE) of the response variable to measure prediction. For the three methods, the RMSEs are calculated as 1.077 (P.Lasso), 1.095 (S.Lasso) and 1.005 (CoFu). In addition, for each gene, we compute the proportion of being identified in 100 resamplings, which has been referred to as the Observed Occurrence Index (OOI) in the literature, to measure the stability of this gene. The highest 20 OOIs are shown in Figure \ref{fig:OOI} (Appendix \ref{sec:table}). CoFu has OOIs (0.823) higher than P.Lasso (0.782) and S.Lasso (0.711).

\subsection{Analysis of lung cancer data}

In TCGA, there are two lung cancer datasets, on Lung Adenocarcinoma (LUAD) and Lung Squamous Cell Carcinoma (LUSC), respectively. In the literature, they have been separately analyzed \cite{cancer2014comprehensive,cancer2012comprehensive}, and differences have been acknowledged \cite{sun2017bioinformatics}. However, as they are both non-small cell lung carcinomas, certain commonality is expected. For both LUAD and LUSC, the inclusion criteria are: (1) no neo-adjuvant therapy before tumor sample collection, (2) in stage I of the AJCC pathologic tumor stage measurement, and (3) with FEV1 (forced expiratory volume in 1sec,  prebroncholiator) and gene expressions measured. More details on sample selection are provided in Figure \ref{fig:flowchart} (Appendix \ref{sec:table}). In this analysis, the response variable is FEV1, a critical measure of lung function. The final sample sizes are 142 (LUAD) and 89 (LUSC), respectively. For gene expressions, we conduct a similar processing as described above. Specifically, 20,531 gene expressions are initially available for analysis. Matching with the KEGG pathway information leads to 4,243 genes. The p-value based marginal screening further reduces the number of gene expressions to 901. Using the same approach as described above, 41 communities are constructed.

\begin{figure}[h] %
   \includegraphics[width=7.5in]{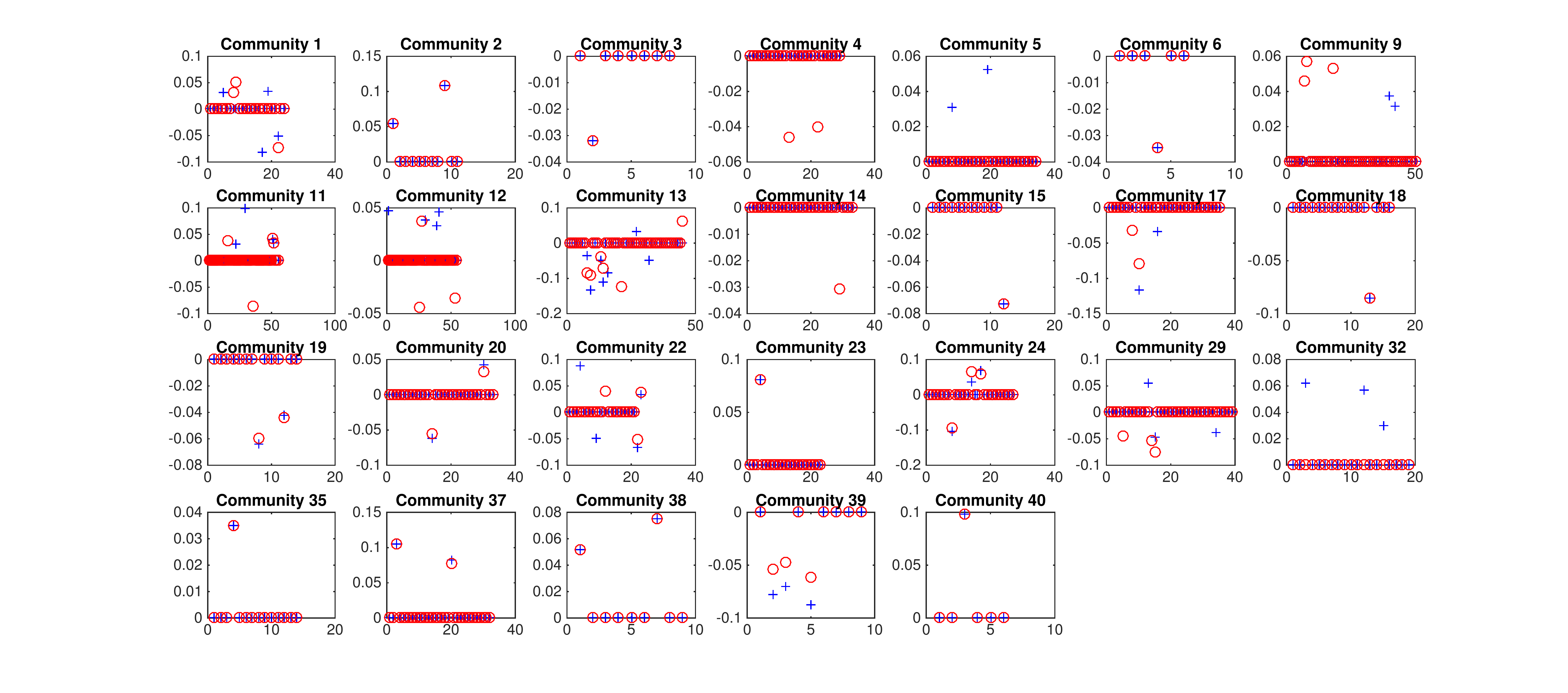}
   \caption{Analysis of the TCGA lung cancer data. Blue crosses correspond to LUAD, and red circles to LUSC.}
   \label{fig:Lung}
\end{figure}

The analysis results of CoFu are summarized in Figure \ref{fig:Lung}. A total of 26 communities, with 54 genes, are identified as associated with the response. Among them, 13 communities behave the same for both LUAD and LUSC. Figure \ref{fig:Lung} suggests that some communities, for example 9 and 13, behave significantly differently for LUAD and LUSC.
Literature review again suggests that the findings are biologically sensible. Specifically, among the identified genes, CCNG2 has the strongest signal for both LUAD and LUSC. CCNG2 encodes protein cyclin G2, which has been shown to be a tumor suppressor in several studies \cite{Wang2016, Yao2017}. A recent study shows that lung cancer patients with higher CCNG2 expressions had longer overall survival. Another gene that has a strong signal in both cancers is BRI3BP, BRI3-binding protein. Over expressions of BRI3BP are found to promote drug induced apoptosis via cross talking between mitochondria and endoplasmic reticulum (ER). We also identify 24 genes that are associated with the response in LUAD but not LUSC. Among them, Chemokine-like receptor 1 (CMKLR1) has the strongest signal. CMLKLR1 is a transmembrane multifunctional receptor and found to play an important role in inflammatory. One genetic variant of CMLKLR1, rs1878022, is found to be significantly associated with poorer survival in advanced stage non-small cell lung cancer \cite{wu2011genome}. Another gene that we find to be associated with lung function in LUAD only is DLGAP5, DLG associated protein 5. DLGAP5 is a mitotic spindle protein and involved in mitosis processes. A study by Schneider et al.\cite{Schneider2017} shows that DLGAP5 expression is higher in lung tumor tissues. Lung cancer patients with the overexpression of DLGAP5 tend to have poorer survival. Genes that have strong signals in LUSC but not LUAD include CALHM2, BTBD3, CLPP, and others. CALHM2 encodes protein calcium homeostasis modulator family member 2, which plays a critical role in the modulation of neural activity via ATP-releasing channel. The BTB domain-containing 3 (BTBD3) gene is found to be upregulated in Hepatocellular carcinoma tissues \cite{Xiao2017}. The genetic variant of BTBD3 is also found to be significantly associated with survival in non-small cell lung cancer patients \cite{Wu2010}. CLPP, caseinolytic mitochondrial matrix peptidase proteolytic subunit, belongs to the peptidase family S14. Its function is to hydrolyze proteins into small peptides in the mitochondria matrix. Increased protein expression is found in type I endometrial cancer patients \cite{Cormio2017}.

The alternative analysis results are presented in Figures \ref{fig:lasso_Lung} (S.Lasso) and \ref{fig:single_Lung} (P.Lasso) in Appendix \ref{sec:table}. S.Lasso identifies 27 communities with 44 genes. No commonality is identified. P.Lasso identifies 31 communities with 44 genes, all of which behave the same for the two cancers. In terms of prediction, the CoFu RMSE is 0.917, lower than S.Lasso (0.999) and P.Lasso (1.023). The OOI results are represented in Figure \ref{fig:OOI} (Appendix \ref{sec:table}). CoFu has the highest OOIs among the three methods.

\subsection{Simulation}
It has been recognized in some studies that simulated data may be ``simpler'' than real data. Here we conduct an additional set of simulation based on the SKCM data analyzed above. Specifically, the observed gene expression data and community structure are used in simulation. The structure of important covariate effects is the same as described above in Section \ref{sec:sim}. The identification results for community and individual effects are summarized in Table \ref{tab:realdata} (Appendix \ref{sec:table}). Although there are some small numerical differences, the observed patterns are similar to those in Section \ref{sec:sim}, providing a strong support to the effectiveness of the proposed method.

\section{Discussion}
In the literature, although the commonality and difference of ``related'' cancers have been noted, effective analysis methods are still lacking. This study fills this knowledge gap by developing a novel community fusion method. The CoFu method has an intuitive formulation and can be effectively realized. Although sharing some similar spirits with the existing fused and contrasted penalization methods, it also has significant advancements by conducting the integrative analysis of multiple datasets, promoting commonality as opposed to similarity, and accommodating the community structure among genes. Numerical studies have demonstrated its superiority over the direct competitors. We have mostly described the proposed method under the linear regression model. As described in Appendix \ref{sec:logit} as well as in the simulation, the proposed method can be extended to the Logit model, a representative of generalized linear models. A closer examination of the penalty function and computation suggests that the CoFu method can be potentially coupled with others, for example prognosis, outcomes and models. More complicated penalties can take the place of Lasso. For example, when it is desirable to accommodate network adjacency, Laplacian penalties can be further imposed. The group Lasso type fusion penalty can also be replaced by other group penalties.
In the first data analysis, the ``stratification'' variable is cancer stage, which has a sound biological basis. It should be noted that, as demonstrated in simulation, the proposed method can accommodate different degrees of commonality/difference. As such, the choice of the stratification variable is not critical. In fact, the proposed method can be applied to ``test'' whether the response-omics relationships are the same with respect to a specific stratification variable.

Limitations of this study may include a lack of theoretical investigation and deeper bioinformatics analysis of the data analysis results. 
Such pursuit will be deferred to future research. We will also defer possible extensions as discussed above to future research.

\section*{Acknowledgments}
We thank the associate editor and two reviewers for careful review and insightful comments, which have led to a significant improvement of the article. We thank Pan Zhang for sharing computer code, which helps data generation in simulation. 
This study was supported by National Natural Science Foundation of China (11605288, 71771211), MOE (Ministry of Education in China) Project of Humanities and Social Sciences (16YJCZH088), Fund for building world-class universities (disciplines) of Renmin University of China, Bureau of Statistics of China (2016LD01), and National Institute of Health (CA204120, CA216017, CA121974).

\subsection*{Conflict of interest}

The authors declare no potential conflict of interests.

%\bibliography{refs}%

\clearpage

\appendix

%\section{Figures\label{app1}}
\section{Estimation under generalized linear models }
\label{sec:logit}
Here we consider extending the CoFu method to the generalized linear models (GLMs). Consider $K$ independent datasets. In dataset $k(=1,\ldots,K)$, denote $\bm y^k$ as the response variable, and $\bm{X}^k$ as the length-$p$ vector of covariates. There are $n_k$ i.i.d observations in dataset $k$. The response variable $\bm y^k$ is generated from a distribution in the exponential family with mean $\bm\mu^k\equiv E(\bm{y}^k)$ dependent on $\bm{X}^k$ via
\begin{equation*}
g(\bm \mu^k)=\bm{X}^k\bm{\beta}^k,
\end{equation*} 
where $g(\cdot)$ is the link function. In the Logit model, a representative of GLM, $g(\mu)=\log\frac{\mu}{1-\mu}$. 

The CoFu estimator $\{\hat{\bm\beta}^k: k=1,\ldots, K\}$ is defined as the minimizer of  
\begin{equation}
 -\sum_{k=1}^K\frac{1}{n_k}L_k(\bm{\beta}^k)+\lambda_1\sum_{k=1}^K\parallel\bm{\beta}^k\parallel_1
+\lambda_2\sum_{k=1}^{K-1}\sum_{l=1}^L\parallel \bm{\beta}^k_{(l)}-\bm{\beta}^{k+1}_{(l)}\parallel_2,
\end{equation}
where $L_k(\cdot)$ is the log-likelihood function of dataset $k$. In the Logit model, $L_k(\bm{\beta}^k)=
\sum_{i=1}^{n_k}y_i^k\bm{X}_i^k\bm{\beta}^k-\log[1+\exp(\bm{X}_i^k\bm{\beta}^k)]$.

We compute the minimizer of (A1) by using the ADMM algorithm: 
\begin{eqnarray*}
\bm{\beta}(t+1) &= & arg min_{\bm{\beta}}\left[-\sum_{k=1}^K\frac{1}{n_k}L_k(\bm{\beta}^k)+\frac{\sigma}{2}\Big(\parallel \bm{A}\bm\beta-\bm\eta(t)+\bm u(t)/\sigma\parallel_2^2+\parallel\bm\beta-\bm\delta(t)+\bm v(t)/\sigma\parallel_2^2\Big)\right], \\
\bm\delta(t+1) &= & \mathrm{ST}_{\lambda_1/\sigma}[\bm\beta(t+1)+\bm v(t)/\sigma],\\
\bm \eta^k_{(l)}(t+1)  &= & \Big( 1-\frac{\lambda_2}{\sigma\parallel \bm\beta^k_{(l)}(t+1)-\bm\beta^{k+1}_{(l)}(t+1)+\bm u^k_{(l)}(t)/\sigma\parallel_2} \Big)_{+}\times [\bm \beta^k_{(l)}(t+1)-\bm \beta^{k+1}_{(l)}(t+1)+\bm u^{k}_{(l)}(t)/\sigma],\\
 \bm u(t+1) &= & \bm u(t)+\sigma[\bm A \bm \beta(t+1)-\bm \eta(t+1)],\\
\bm v(t+1)  & = & \bm v(t)+\sigma[\bm \beta(t+1)-\bm \delta(t+1)].
\end{eqnarray*}
These update equations are highly similar to that in the linear regression model. The $\bm{\beta}$ update involves solving a non-linear equation and can be efficiently realized with the L-BFGS algorithm\cite{Nocedal1980}, a limited memory version of the BFGS algorithm.

\section{Additional tables and figures}
\label{sec:table}
\begin{figure}[h] %
   \centering
   \includegraphics[width=5in]{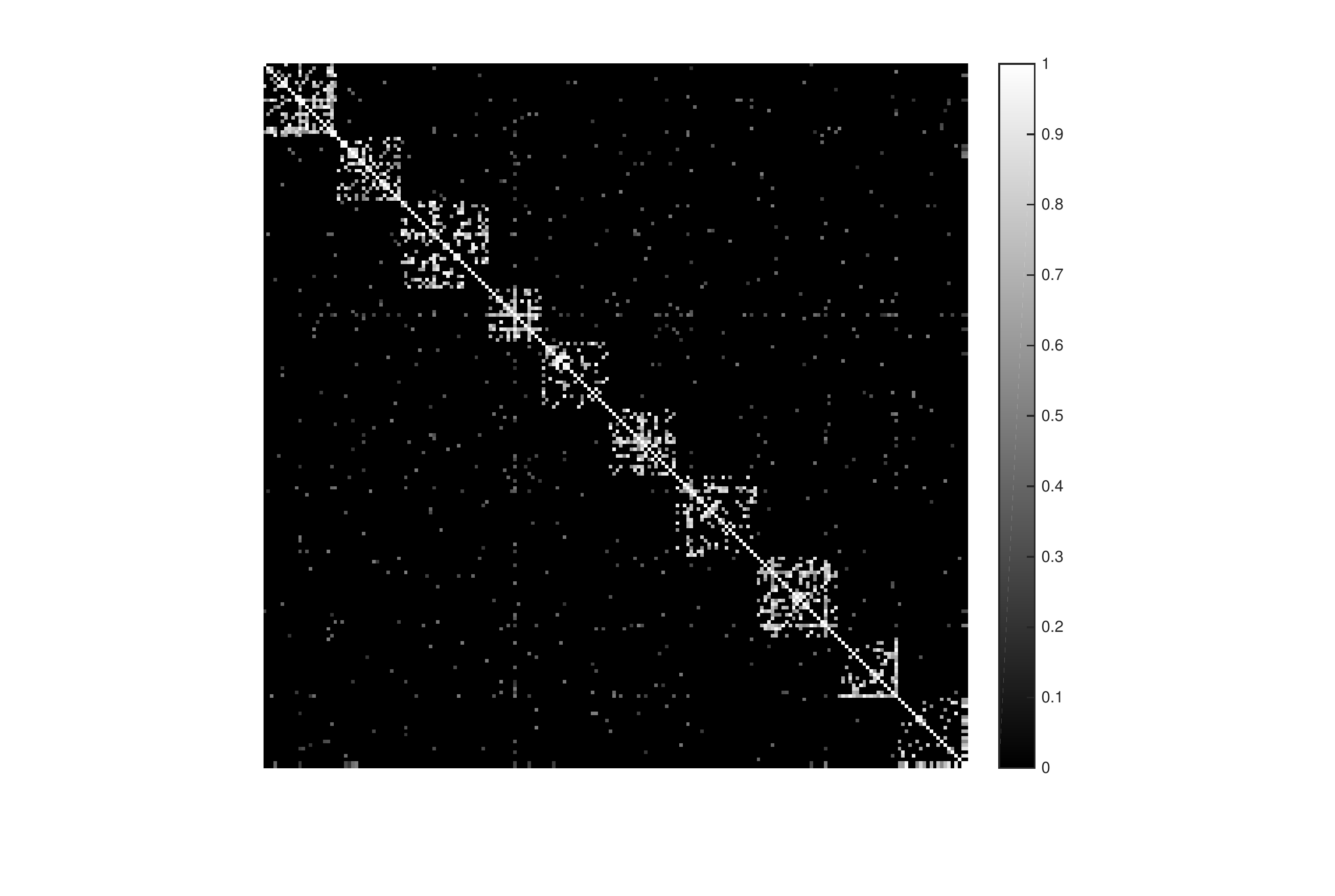}
   \caption{Greyscale image of a structured correlation matrix. The lighter the color is, the stronger the correlation is.\label{fig:sigma}}
\end{figure}

\clearpage
\begin{table}[htbp]
   \centering
   \caption{Simulation under the LR model: mean(sd) of AUC for effect identification. Nonzero coefficients are randomly drawn from $\mathcal{U}[0.2,1]$.} % requires the topcapt package
   \begin{tabular}{lcccccc} % Column formatting, @{} suppresses leading/trailing space
      \toprule
      %\multicolumn{2}{c}{correlation} \\
      & & $r=100$ & &  & $r=150$\\
      \cmidrule(r){2-4} \cmidrule(l){5-7}% Partial rule. (r) trims the line a little bit on the right; (l) & (lr) also possible
         & structured &unstructured & independence  &structured &unstructured & independence \\
      \midrule
       \multicolumn{3}{l}{$(\rho_a,\rho_h,\rho_n)=(0.1,0,0.9)$} \\
        P.Lasso &0.605(0.018)&0.57(0.01)&0.612(0.017)&0.578(0.016)&0.576(0.02)&0.595(0.019)\\
       S.Lasso  &0.746(0.012)&0.795(0.008)&0.745(0.01)&0.677(0.015)&0.75(0.011)&0.675(0.017)\\
       CoFu &0.761(0.011)&0.848(0.01)&0.757(0.012)&0.695(0.013)&0.746(0.015)&0.688(0.012)\\
        \multicolumn{3}{l}{$(\rho_a,\rho_h,\rho_n)=(0.1,0.9,0)$} \\
        P.Lasso &0.744(0.021)&0.698(0.014)&0.745(0.015)&0.659(0.017)&0.646(0.013)&0.661(0.014)\\
       S.Lasso  &0.757(0.014)&0.807(0.013)&0.752(0.014)&0.683(0.017)&0.756(0.011)&0.676(0.015)\\
       CoFu &0.839(0.013)&0.848(0.007)&0.844(0.012)&0.763(0.013)&0.795(0.009)&0.744(0.01)\\
      \multicolumn{3}{l}{$(\rho_a,\rho_h,\rho_n)=(0.2,0.6,0.2)$} \\
       P.Lasso & 0.721(0.023) &0.827(0.012) &0.713(0.019) &0.667(0.015) &0.647(0.015) &0.658(0.016)\\
       S.Lasso  & 0.757(0.019) & 0.806(0.009) & 0.756(0.011) & 0.689(0.01) &0.763(0.017) &0.677(0.013)\\
       CoFu &0.824(0.015) &0.882(0.008) &0.82(0.013) & 0.751(0.012) &0.786(0.014) &0.738(0.014)\\
       \multicolumn{3}{l}{$(\rho_a,\rho_h,\rho_n)=(0.4,0.1,0.5)$} \\
       P.Lasso & 0.723(0.01) &0.713(0.013) &0.706(0.014) &0.675(0.014) &0.666(0.014) &0.673(0.018)\\
       S.Lasso  & 0.752(0.009) & 0.801(0.012) & 0.748(0.008) &0.681(0.011) &0.757(0.012) &0.668(0.012)\\
       CoFu &0.82(0.009) &0.843(0.011) &0.804(0.014) &0.74(0.014) &0.779(0.01) &0.733(0.013)\\
        \multicolumn{3}{l}{$(\rho_a,\rho_h,\rho_n)=(0.5,0.5,0)$} \\
      P.Lasso & 0.845(0.02) &0.827(0.012) &0.838(0.018) &0.787(0.019) &0.777(0.015) &0.781(0.021)\\
       S.Lasso  & 0.76(0.015) &0.806 (0.009)& 0.753(0.013) & 0.691(0.011) &0.759(0.016) &0.682(0.013)\\
       CoFu &0.88(0.014) &0.882(0.008) &0.876(0.01) &0.805(0.01) &0.825(0.012)&0.806(0.008) \\
       \multicolumn{3}{l}{$(\rho_a,\rho_h,\rho_n)=(0.6,0.2,0.2)$} \\
        P.Lasso & 0.852(0.012) &0.819(0.015) &0.838(0.016) &0.79(0.015) &0.765(0.021) &0.788(0.017)\\
       S.Lasso  & 0.755(0.013) & 0.803(0.011) & 0.753(0.011) &0.684(0.022) &0.755(0.014) &0.677(0.018)\\
       CoFu &0.883(0.012) &0.877(0.009) &0.866(0.01) &0.803(0.016) &0.811(0.013) &0.802(0.016)\\
         \multicolumn{3}{l}{$(\rho_a,\rho_h,\rho_n)=(0.9,0,0.1)$} \\
        P.Lasso &0.955(0.015)&0.889(0.014)&0.954(0.014)&0.886(0.016)&0.853(0.017)&0.895(0.014)\\
       S.Lasso  &0.748(0.016)&0.795(0.012)&0.741(0.009)&0.696(0.016)&0.747(0.013)&0.687(0.016)\\
       CoFu &0.967(0.013)&0.93(0.01)&0.964(0.012)&0.92(0.014)&0.887(0.011)&0.92(0.012)\\

      \bottomrule
   \end{tabular}
   \label{tab:var2}
\end{table}

\clearpage
\begin{table}[htbp]
   \centering
   \caption{Simulation under the LR model: mean(sd) of AUC for community differentiation. Nonzero coefficients are randomly drawn from $\mathcal{U}[0.2,1]$.} % requires the topcapt package
   \begin{tabular}{lcccccc} % Column formatting, @{} suppresses leading/trailing space
      \toprule
      %\multicolumn{2}{c}{correlation} \\
      & & $r=100$ & &  & $r=150$\\
      \cmidrule(r){2-4} \cmidrule(l){5-7}% Partial rule. (r) trims the line a little bit on the right; (l) & (lr) also possible
         & structured &unstructured & independence  &structured &unstructured & independence \\
      \midrule
        \multicolumn{3}{l}{$(\rho_a,\rho_h,\rho_n)=(0.1,0,0.9)$} \\
       S.Lasso  &0.524(0.066)&0.551(0.065)&0.554(0.082)&0.563(0.052)&0.591(0.072)&0.537(0.045)\\
       CoFu &0.774(0.042)&0.723(0.039)&0.795(0.038)&0.714(0.045)&0.711(0.048)&0.724(0.037)\\
       \multicolumn{3}{l}{$(\rho_a,\rho_h,\rho_n)=(0.1,0.9,0.1)$} \\
       S.Lasso  &0.626(0.062)&0.53(0.075)&0.595(0.088)&0.627(0.067)&0.652(0.082)&0.675(0.042)\\
       CoFu &0.801(0.045)&0.722(0.032)&0.735(0.034)&0.76(0.041)&0.781(0.035)&0.724(0.052)\\
      \multicolumn{3}{l}{$(\rho_a,\rho_h,\rho_n)=(0.2,0.6,0.2)$} \\
       S.Lasso  & 0.573(0.082) & 0.54(0.105) & 0.549(0.074) & 0.477(0.06) &0.515(0.088) &0.488(0.037)\\
       CoFu   & 0.742(0.028) &0.698(0.042) &0.73(0.04) & 0.711(0.035) &0.734(0.031) &0.711(0.033)\\
       \multicolumn{3}{l}{$(\rho_a,\rho_h,\rho_n)=(0.4,0.1,0.5)$} \\
       S.Lasso  & 0.534(0.072) & 0.552(0.086) & 0.567(0.041) &0.537(0.058) &0.542(0.055) &0.544(0.066)\\
       CoFu &0.785(0.021) &0.738(0.046) &0.777(0.036) &0.727(0.033) &0.731(0.041) &0.715(0.033)\\
       \multicolumn{3}{l}{$(\rho_a,\rho_h,\rho_n)=(0.5,0.5,0)$} \\
       S.Lasso  & 0.542(0.057) & 0.487(0.059) & 0.546(0.061) & 0.515(0.088) &0.492(0.124) &0.513(0.084)\\
       CoFu  &0.812(0.037) &0.754(0.023) &0.81(0.036) & 0.711(0.043) &0.705(0.053) &0.703(0.034)\\
       \multicolumn{3}{l}{$(\rho_a,\rho_h,\rho_n)=(0.6,0.2,0.2)$} \\
       S.Lasso  & 0.5(0.074) & 0.49(0.072) & 0.539(0.103) &0.47(0.075) &0.529(0.08) &0.504(0.068)\\
       CoFu &0.791(0.046) &0.742(0.029) &0.802(0.046) &0.723(0.049) &0.721(0.036) &0.713(0.055)\\
        \multicolumn{3}{l}{$(\rho_a,\rho_h,\rho_n)=(0.9,0,0.1)$} \\
       S.Lasso  &0.481(0.062)&0.45(0.088)&0.453(0.092)&0.52(0.072)&0.478(0.082)&0.444(0.062)\\
       CoFu &0.975(0.031)&0.968(0.033)&0.975(.034)&0.927(0.035)&0.91(0.045)&0.911(0.031)\\
      \bottomrule
   \end{tabular}
   %\caption{Remember, \emph{never} use vertical lines in tables.}
   \label{tab:group2}
\end{table}
%\end{sidewaystable}

\clearpage
\begin{table}[htbp]
   \centering
   \caption{
   Simulation under the LR model: mean(sd) of AUC for effect identification. Nonzero coefficients in dataset 2 are drawn from $\mathcal{U}[0.4,0.7]$, and those specific to datasets 1 and 3 are drawn from $\mathcal{U}[0.1,0.3]$ and $\mathcal{U}[0.8,1]$, respectively.} % requires the topcapt package
   \begin{tabular}{lcccccc} % Column formatting, @{} suppresses leading/trailing space
      \toprule
      %\multicolumn{2}{c}{correlation} \\
      & & $r=100$ & &  & $r=150$\\
      \cmidrule(r){2-4} \cmidrule(l){5-7}% Partial rule. (r) trims the line a little bit on the right; (l) & (lr) also possible
         & structured &unstructured & independence  &structured &unstructured & independence \\
      \midrule
       \multicolumn{3}{l}{$(\rho_a,\rho_h,\rho_n)=(0.1,0,0.9)$} \\
        P.Lasso &0.598(0.02)&0.549(0.022)&0.591(0.012)&0.559(0.02)&0.546(0.014)&0.556(0.021)\\
       S.Lasso  & 0.757(0.011)&0.812(0.012)&0.738(0.015)&0.675(0.021)&0.757(0.011)&0.666(0.015)\\
       CoFu &0.759(0.012)&0.805(0.009)&0.744(0.012)&0.677(0.018)&0.734(0.012)&0.661(0.018)\\
        \multicolumn{3}{l}{$(\rho_a,\rho_h,\rho_n)=(0.1,0.9,0)$} \\
        P.Lasso &0.721(0.02)&0.684(0.022)&0.744(0.012)&0.675(0.016)&0.621(0.012)&0.667(0.015)\\
       S.Lasso  & 0.768(0.013)&0.832(0.012)&0.761(0.015)&0.689(0.018)&0.755(0.012)&0.68(0.014)\\
       CoFu &0.811(0.012)&0.846(0.01)&0.813(0.013)&0.761(0.016)&0.786(0.01)&0.754(0.017)\\
      \multicolumn{3}{l}{$(\rho_a,\rho_h,\rho_n)=(0.2,0.6,0.2)$} \\
       P.Lasso & 0.724(0.023) &0.678(0.017) &0.726(0.009) &0.66(0.022) &0.624(0.01) &0.662(0.022)\\
       S.Lasso  & 0.761(0.01) & 0.819(0.014) & 0.756(0.007) & 0.69(0.014) &0.764(0.013) &0.681(0.012)\\
       CoFu   & 0.823(0.015) &0.853(0.011) &0.82(0.012) & 0.754(0.011) &0.785(0.014) &0.741(0.01)\\
       \multicolumn{3}{l}{$(\rho_a,\rho_h,\rho_n)=(0.4,0.1,0.5)$} \\
       P.Lasso & 0.712(0.018) &0.685(0.027) &0.708(0.014) &0.672(0.012) &0.626(0.018) &0.67(0.012)\\
       S.Lasso  & 0.751(0.012) & 0.806(0.009) & 0.731(0.022) &0.681(0.013) &0.758(0.009) &0.673(0.013)\\
       CoFu &0.809(0.013) &0.847(0.007) &0.804(0.014) &0.815(0.02) &0.786(0.008) &0.734(0.022)\\
        \multicolumn{3}{l}{$(\rho_a,\rho_h,\rho_n)=(0.5,0.5,0)$} \\
      P.Lasso & 0.868(0.016) &0.845(0.012) &0.868(0.013) &0.799(0.021) &0.766(0.021) &0.796(0.019)\\
       S.Lasso  & 0.769(0.014) & 0.82(0.006) & 0.757(0.019) & 0.693(0.011) &0.768(0.013) &0.684(0.009)\\
       CoFu  &0.889(0.014) &0.884(0.006) &0.885(0.013) & 0.816(0.014) &0.829(0.01) &0.809(0.016)\\
       \multicolumn{3}{l}{$(\rho_a,\rho_h,\rho_n)=(0.6,0.2,0.2)$} \\
        P.Lasso & 0.861(0.016) &0.827(0.019) &0.858(0.014) &0.807(0.018) &0.751(0.015) &0.805(0.024)\\
       S.Lasso  & 0.764(0.012) & 0.814(0.009) & 0.751(0.013) &0.686(0.023) &0.761(0.01) &0.677(0.018)\\
       CoFu &0.878(0.008) &0.877(0.007) &0.872(0.01) &0.815(0.02) &0.817(0.011) &0.813(0.02)\\
        \multicolumn{3}{l}{$(\rho_a,\rho_h,\rho_n)=(0.9,0,0.1)$} \\
        P.Lasso &0.878(0.02)&0.889(0.022)&0.861(0.012)&0.909(0.019)&0.855(0.019)&0.913(0.016)\\
       S.Lasso  &0.761(0.013)&0.813(0.008)&0.749(0.018)&0.68(0.018)&0.75(0.012)&0.672(0.015)\\
       CoFu &0.857(0.012)&0.883(0.009)&0.848(0.012)&0.932(0.017)&0.899(0.013)&0.93(0.021)\\
      \bottomrule
   \end{tabular}
   %\caption{Remember, \emph{never} use vertical lines in tables.}
   \label{tab:var3}
\end{table}

%\begin{sidewaystable}\small
\begin{table}[htbp]
   \centering
   \caption{Simulation under the LR model: mean(sd) of AUC for community differentiation. Nonzero coefficients in dataset 2 are drawn from $\mathcal{U}[0.4,0.7]$, and those specific to datasets 1 and 3 are drawn from $\mathcal{U}[0.1,0.3]$ and $\mathcal{U}[0.8,1]$, respectively.} % requires the topcapt package
   \begin{tabular}{lcccccc} % Column formatting, @{} suppresses leading/trailing space
      \toprule
      %\multicolumn{2}{c}{correlation} \\
      & & $r=100$ & &  & $r=150$\\
      \cmidrule(r){2-4} \cmidrule(l){5-7}% Partial rule. (r) trims the line a little bit on the right; (l) & (lr) also possible
         & structured &unstructured & independence  &structured &unstructured & independence \\
      \midrule
       \multicolumn{3}{l}{$(\rho_a,\rho_h,\rho_n)=(0.1,0,0.9)$} \\
       S.Lasso  &0.598(0.066)&0.609(0.086)&0.553(0.062)&0.578(0.068)&0.515(0.082)&0.462(0.088)\\
       CoFu &0.793(0.036)&0.782(0.052)&0.787(0.042)&0.824(0.038)&0.84(0.038)&0.853(0.035)\\
        \multicolumn{3}{l}{$(\rho_a,\rho_h,\rho_n)=(0.1,0.9,0)$} \\
       S.Lasso  &0.449(0.055)&0.441(0.082)&0.489(0.056)&0.658(0.061)&0.629(0.082)&0.6(0.058)\\
       CoFu &0.765(0.035)&0.754(0.031)&0.789(0.040)&0.79(0.037)&0.737(0.042)&0.733(0.035)\\
      \multicolumn{3}{l}{$(\rho_a,\rho_h,\rho_n)=(0.2,0.6,0.2)$} \\
       S.Lasso  & 0.578(0.083) & 0.516(0.093) & 0.574(0.078) & 0.499(0.077) &0.544(0.1) &0.506(0.094)\\
       CoFu   & 0.733(0.042) &0.713(0.057) &0.734(0.037) & 0.711(0.035) &0.734(0.031) &0.711(0.033)\\
       \multicolumn{3}{l}{$(\rho_a,\rho_h,\rho_n)=(0.4,0.1,0.5)$} \\
       S.Lasso  & 0.553(0.038) & 0.546(0.071) & 0.558(0.046) &0.516(0.063) &0.516(0.06) &0.541(0.061)\\
       CoFu &0.755(0.041) &0.754(0.044) &0.753(0.027) &0.724(0.034) &0.719(0.047) &0.695(0.029)\\
         \multicolumn{3}{l}{$(\rho_a,\rho_h,\rho_n)=(0.5,0.5,0)$} \\
       S.Lasso  & 0.53(0.072) & 0.48(0.063) & 0.549(0.073) & 0.474(0.063) &0.515(0.094) &0.502(0.052)\\
       CoFu  &0.811(0.039) &0.769(0.026) &0.818(0.047) & 0.698(0.04) &0.692(0.059) &0.701(0.028)\\
       \multicolumn{3}{l}{$(\rho_a,\rho_h,\rho_n)=(0.6,0.2,0.2)$} \\
       S.Lasso  & 0.495(0.071) & 0.482(0.082) & 0.476(0.073) &0.476(0.065) &0.51(0.073) &0.504(0.075)\\
       CoFu &0.787(0.037) &0.762(0.031) &0.747(0.031) &0.724(0.029) &0.702(0.054) &0.721(0.038)\\
      \multicolumn{3}{l}{$(\rho_a,\rho_h,\rho_n)=(0.9,0,0.1)$} \\
       S.Lasso  &0.529(0.062)&0.539(0.067)&0.543(0.071)&0.449(0.068)&0.519(0.082)&0.584(0.082)\\
       CoFu &0.975(0.036)&0.968(0.036)&0.975(0.035)&0.899(0.03)&0.908(0.042)&0.908(0.035))\\
      \bottomrule
   \end{tabular}
   %\caption{Remember, \emph{never} use vertical lines in tables.}
   \label{tab:group3}
\end{table}
%\end{sidewaystable}

\clearpage
\begin{table}[htbp]
   \centering
   \caption{Simulation under the LR model: summary statistics of results obtained by using V-fold (V=5) CV. Nonzero coefficients are equal to 0.5. In each cell, mean(sd).} % requires the topcapt package
   \begin{tabular}{lcccccc} % Column formatting, @{} suppresses leading/trailing space
      \toprule
      & \multicolumn{2}{c}{Individual effect} & \multicolumn{2}{c}{Community}   \\
      \cmidrule(r){2-3} \cmidrule(l){4-5} % Partial rule. (r) trims the line a little bit on the right; (l) & (lr) also possible
         & TPR &FPR &  TPR  & FPR &  ERMSE & PRMSE \\
      \midrule
      \multicolumn{5}{l}{$(\rho_a,\rho_h,\rho_n)=(0.1,0,0.9)$} \\
       P.Lasso & 0.176 (0.054) &0.05 (0.025) & 1 (0)  & 1 (0) & 0.154 (0.003) & 4.181 (0.414) \\
       S.Lasso & 0.238 (0.07)  & 0.041 (0.015) & 0.179 (0.118) & 0.022 (0.032) &0.151 (0.002) & 3.335 (0.533) \\
       CoFu     & 0.625 (0.033) &0.21 (0.026)& 0.579 (0.158) & 0.072 (0.059) & 0.144 (0.001) & 0.838 (0.109) \\
      \multicolumn{5}{l}{$(\rho_a,\rho_h,\rho_n)=(0.1,0.9,0)$} \\
        P.Lasso & 0.449(0.06) &0.081(0.018) & 1 (0)  & 1 (0) & 0.132 (0.002) & 2.937(0.167) \\
       S.Lasso & 0.291(0.074)& 0.036(0.013) & 0.23(0.157) & 0.065(0.09) &0.135(0.003) & 2.88 (0.463) \\
       CoFu&   0.732(0.024) &0.211(0.017)& 0.67(0.106) & 0.205(0.037) & 0.122(0.003) & 0.685(0.047) \\
       \multicolumn{5}{l}{$(\rho_a,\rho_h,\rho_n)=(0.2,0.6,0.2)$} \\
       P.Lasso & 0.496 (0.071) &0.104 (0.035) & 1 (0)  & 1 (0) & 0.133 (0.002) & 2.868 (0.441) \\
       S.Lasso &0.319 (0.069) & 0.045 (0.016) & 0.177 (0.162) & 0.016 (0.019) & 0.138 (0.002) & 2.601 (0.63) \\
       CoFu&   0.77 (0.029) & 0.214 (0.017) & 0.685 (0.099) & 0.16 (0.076) & 0.122 (0.003) & 0.69 (0.057) \\
       \multicolumn{5}{l}{$(\rho_a,\rho_h,\rho_n)=(0.4,0.1,0.5)$} \\
       P.Lasso & 0.523 (0.043) &0.111 (0.025) & 1 (0)  & 1 (0) & 0.135 (0.002) & 2.750 (0.251) \\
       S.Lasso & 0.279 (0.06)  & 0.039 (0.014) & 0.142 (0.071) & 0.012 (0.026) & 0.146 (0.002) & 2.926 (0.554) \\
       CoFu&   0.76 (0.032) & 0.206 (0.011) & 0.579 (0.069) & 0.104 (0.051) & 0.126 (0.003) & 0.703 (0.039) \\
       \multicolumn{5}{l}{$(\rho_a,\rho_h,\rho_n)=(0.5,0.5,0)$} \\
       P.Lasso & 0.772 (0.03) &0.12 (0.019) & 1 (0)  & 1 (0) & 0.111 (0.004) & 1.945 (0.203) \\
       S.Lasso & 0.316 (0.069)& 0.047 (0.012) & 0.107 (0.082) & 0.029 (0.051) & 0.142 (0.003) & 2.697 (0.557) \\
       CoFu&   0.909 (0.022) &0.168 (0.013)& 0.769 (0.08) & 0.138 (0.057) & 0.1 (0.004) & 0.669 (0.027) \\
       \multicolumn{5}{l}{$(\rho_a,\rho_h,\rho_n)=(0.6,0.2,0.2)$} \\
       P.Lasso & 0.729 (0.032) &0.141 (0.028) & 1 (0)  & 1 (0) & 0.118 (0.004) & 1.9 (0.294) \\
       S.Lasso & 0.28 (0.11)& 0.041 (0.027) & 0.158 (0.154) & 0.029 (0.057) & 0.146 (0.003) & 2.879 (1.19) \\
       CoFu&   0.87 (0.02) &0.182 (0.009)& 0.797 (0.088) & 0.135 (0.068) & 0.107 (0.004) & 0.655 (0.024) \\
       \multicolumn{5}{l}{$(\rho_a,\rho_h,\rho_n)=(0.9,0,0.1)$} \\
       P.Lasso & 0.915 (0.037) &0.132 (0.034) & 1 (0)  & 1 (0) & 0.105 (0.004) & 1.685 (0.308) \\
       S.Lasso & 0.275 (0.077)& 0.044 (0.02) & 0.12 (0.098) & 0.03 (0.048) & 0.149 (0.002) & 2.996 (0.721) \\
       CoFu&   0.925 (0.025) & 0.161 (0.02)& 0.925 (0.09) & 0.05 (0.071) & 0.095 (0.006) & 0.65 (0.036) \\

      \bottomrule
   \end{tabular}
   %\caption{Remember, \emph{never} use vertical lines in tables.}
   \label{tab:CV1}
\end{table}

\clearpage
\begin{table}[htbp]
   \centering
   \caption{Simulation under the LR model: summary statistics of results obtained by using V-fold (V=5) CV. Nonzero coefficients are randomly drawn from $\mathcal{U}[0.2,1]$. In each cell, mean(sd).} % requires the topcapt package
   \begin{tabular}{lcccccc} % Column formatting, @{} suppresses leading/trailing space
      \toprule
      & \multicolumn{2}{c}{Individual effect} & \multicolumn{2}{c}{Community}  \\
      \cmidrule(r){2-3} \cmidrule(l){4-5} % Partial rule. (r) trims the line a little bit on the right; (l) & (lr) also possible
         & TPR &FPR &  TPR  & FPR &  ERMSE & RMSPE \\
      \midrule
      \multicolumn{5}{l}{$(\rho_a,\rho_h,\rho_n)=(0.1,0,0.9)$} \\
       P.Lasso & 0.232 (0.047) &0.08 (0.029) & 1 (0)  & 1 (0) & 0.197 (0.001) & 4.875 (0.599) \\
       S.Lasso & 0.296 (0.065)  & 0.044 (0.015) & 0.121 (0.089) & 0.011 (0.019) & 0.185 (0.007) & 3.657 (0.55) \\
       CoFu     & 0.617 (0.044) &0.228 (0.02)& 0.583 (0.138) & 0.058 (0.021) & 0.174 (0.004) & 0.913 (0.112) \\
      \multicolumn{5}{l}{$(\rho_a,\rho_h,\rho_n)=(0.1,0.9,0)$} \\
        P.Lasso & 0.497 (0.054) &0.117 (0.027) & 1 (0)  & 1 (0) & 0.162 (0.003) & 3.388 (0.4) \\
       S.Lasso & 0.413 (0.061) & 0.059 (0.015) & 0.057 (0.056) & 0.017 (0.03) & 0.159 (0.006) & 2.459 (0.673) \\
       CoFu&   0.748 (0.033) &0.202 (0.011)& 0.642 (0.084) & 0.114 (0.031) & 0.138 (0.003) & 0.766 (0.028) \\
       \multicolumn{5}{l}{$(\rho_a,\rho_h,\rho_n)=(0.2,0.6,0.2)$} \\
       P.Lasso & 0.535 (0.052) &0.121 (0.038) & 1 (0)  & 1 (0) & 0.162 (0.004) & 3.249 (0.436) \\
       S.Lasso &0.353 (0.071) & 0.051 (0.015) & 0.107 (0.077) & 0.025 (0.021) & 0.173 (0.006) & 2.97 (0.652) \\
       CoFu&   0.759 (0.023) & 0.205 (0.01) & 0.629 (0.121) & 0.125 (0.038) & 0.142 (0.006) & 0.771 (0.036) \\
       \multicolumn{5}{l}{$(\rho_a,\rho_h,\rho_n)=(0.4,0.1,0.5)$} \\
       P.Lasso & 0.457 (0.047) &0.097 (0.023) & 1 (0)  & 1 (0) & 0.162 (0.003) & 3.615 (0.401) \\
       S.Lasso & 0.374 (0.058)  & 0.062 (0.013) & 0.013 (0.033) & 0.038 (0) & 0.172 (0.004) & 2.69 (0.679) \\
       CoFu&   0.725 (0.037) & 0.204 (0.018) & 0.67 (0.096) & 0.096 (0.042) & 0.143 (0.003) & 0.821 (0.056) \\
       \multicolumn{5}{l}{$(\rho_a,\rho_h,\rho_n)=(0.5,0.5,0)$} \\
       P.Lasso & 0.752 (0.05) &0.114 (0.023) & 1 (0)  & 1 (0) & 0.115 (0.004) & 2.114 (0.264) \\
       S.Lasso & 0.427 (0.05)& 0.064 (0.008) & 0.032 (0.03) & 0.016 (0.025) & 0.152 (0.007) & 2.234 (0.583) \\
       CoFu&   0.848 (0.019) &0.139 (0.011)& 0.668 (0.087) & 0.116 (0.054) & 0.097 (0.003) & 0.732 (0.037) \\
       \multicolumn{5}{l}{$(\rho_a,\rho_h,\rho_n)=(0.6,0.2,0.2)$} \\
       P.Lasso & 0.701 (0.038) &0.149 (0.028) & 1 (0)  & 1 (0) & 0.142 (0.005) & 2.748 (0.307) \\
       S.Lasso & 0.376 (0.054)& 0.062 (0.02) & 0.049 (0.029) & 0.012 (0.025) & 0.178 (0.007) & 2.707 (1.072) \\
       CoFu&   0.849 (0.035) &0.205 (0.016) & 0.8 (0.087) & 0.147 (0.064) & 0.126 (0.003) & 0.986 (0.036) \\
       \multicolumn{5}{l}{$(\rho_a,\rho_h,\rho_n)=(0.9,0,0.1)$} \\
       P.Lasso & 0.784 (0.034) &0.126 (0.027) & 1 (0)  & 1 (0) & 0.116 (0.003) & 2.01 (0.267) \\
       S.Lasso & 0.336 (0.092)& 0.056 (0.02) & 0.066 (0.085) & 0.02 (0.047) & 0.181 (0.007) & 3.099 (1.032) \\
       CoFu&   0.886 (0.027) & 0.16 (0.019)& 0.912 (0.048) & 0.133 (0.126) & 0.098 (0.006) & 0.971 (0.027) \\

      \bottomrule
   \end{tabular}
   %\caption{Remember, \emph{never} use vertical lines in tables.}
   \label{tab:CV2}
\end{table}

\clearpage
\begin{table}[htbp]

   \centering
   \caption{
  Simulation under the LR model: summary statistics of results obtained by using V-fold (V=5) CV. Nonzero coefficients in dataset 2 are drawn from $\mathcal{U}[0.4,0.7]$, and those specific to datasets 1 and 3 are drawn from $\mathcal{U}[0.1,0.3]$ and $\mathcal{U}[0.8,1]$, respectively. In each cell, mean(sd).} % requires the topcapt package
   \begin{tabular}{lcccccc} % Column formatting, @{} suppresses leading/trailing space
      \toprule
      & \multicolumn{2}{c}{Individual effect} & \multicolumn{2}{c}{Community}   \\
      \cmidrule(r){2-3} \cmidrule(l){4-5} % Partial rule. (r) trims the line a little bit on the right; (l) & (lr) also possible
         & TPR &FPR &  TPR  & FPR &  ERMSE & PRMSE \\
      \midrule
      \multicolumn{5}{l}{$(\rho_a,\rho_h,\rho_n)=(0.1,0,0.9)$} \\
       P.Lasso & 0.213 (0.083) &0.057 (0.04) & 1 (0)  & 1 (0) & 0.192 (0.002) & 4.807 (1.013) \\
       S.Lasso & 0.253 (0.067)  & 0.025 (0.01) & 0.207 (0.154) & 0.002 (0.01) & 0.181 (0.005) & 3.694 (0.723) \\
       CoFu     & 0.664 (0.044) & 0.186 (0.027)& 0.597 (0.124) & 0.094 (0.004) & 0.168 (0.005) & 0.997 (0.187) \\
      \multicolumn{5}{l}{$(\rho_a,\rho_h,\rho_n)=(0.1,0.9,0)$} \\
        P.Lasso & 0.462 (0.088) &0.101 (0.032) & 1 (0)  & 1 (0) & 0.155 (0.004) & 3.421 (0.466) \\
       S.Lasso & 0.38 (0.075) & 0.051 (0.014) & 0.114 (0.096) & 0.008 (0.013) & 0.149 (0.003) & 2.471 (0.738) \\
       CoFu&   0.759 (0.02) &0.191 (0.013)& 0.612 (0.096) & 0.117 (0.041) & 0.132 (0.003) & 0.772 (0.049) \\
       \multicolumn{5}{l}{$(\rho_a,\rho_h,\rho_n)=(0.2,0.6,0.2)$} \\
       P.Lasso & 0.513 (0.05) &0.124 (0.025) & 1 (0)  & 1 (0) & 0.158 (0.003) & 3.22 (0.269) \\
       S.Lasso &0.363 (0.073) & 0.055 (0.017) & 0.044 (0.057) & 0.009 (0.015) & 0.158 (0.004) & 2.43 (0.801) \\
       CoFu&   0.743 (0.026) & 0.2 (0.013) & 0.63 (0.064) & 0.113 (0.052) & 0.138 (0.004) & 0.765 (0.026) \\
       \multicolumn{5}{l}{$(\rho_a,\rho_h,\rho_n)=(0.4,0.1,0.5)$} \\
       P.Lasso & 0.547 (0.067) & 0.11 (0.028) & 1 (0)  & 1 (0) & 0.166 (0.004) & 3.604 (0.339) \\
       S.Lasso & 0.378 (0.061)  & 0.056 (0.023) & 0.056 (0.044) & 0.004 (0.014) & 0.172 (0.002) & 2.75 (1) \\
       CoFu&   0.836 (0.026) & 0.189 (0.02) & 0.632 (0.075) & 0.108 (0.048) & 0.144 (0.003) & 0.844 (0.059) \\
       \multicolumn{5}{l}{$(\rho_a,\rho_h,\rho_n)=(0.5,0.5,0)$} \\
       P.Lasso & 0.793 (0.038) &0.132 (0.021) & 1 (0)  & 1 (0) & 0.124 (0.004) & 2.175 (0.291) \\
       S.Lasso & 0.316 (0.05)& 0.05 (0.015) & 0.069 (0.066) & 0.017 (0.038) & 0.16 (0.004) & 2.996 (0.755) \\
       CoFu&   0.905 (0.019) &0.16 (0.011)& 0.689 (0.117) & 0.056 (0.026) & 0.106 (0.004) & 0.715 (0.028) \\
       \multicolumn{5}{l}{$(\rho_a,\rho_h,\rho_n)=(0.6,0.2,0.2)$} \\
       P.Lasso & 0.811 (0.04) &0.148 (0.042) & 1 (0)  & 1 (0) & 0.126 (0.005) & 1.971 (0.468) \\
       S.Lasso & 0.306 (0.073)& 0.045 (0.013) & 0.078 (0.061) & 0.021 (0.048) & 0.167 (0.005) & 3.35 (0.793) \\
       CoFu&   0.884 (0.021) &0.193 (0.018) & 0.867 (0.05) & 0.164 (0.117) & 0.112 (0.007) & 0.902 (0.032) \\
       \multicolumn{5}{l}{$(\rho_a,\rho_h,\rho_n)=(0.9,0,0.1)$} \\
       P.Lasso & 0.969 (0.007) &0.104 (0.016) & 1 (0)  & 1 (0) & 0.073 (0.003) & 1.17 (0.107) \\
       S.Lasso & 0.33 (0.048) & 0.053 (0.013) & 0.067 (0.055) & 0 (0) & 0.166 (0.004) & 1.985 (0.59) \\
       CoFu&   0.982 (0.005) & 0.083 (0.025)& 0.998 (0.007) & 0 (0) & 0.054 (0.005) & 0.801 (0.041) \\

      \bottomrule
   \end{tabular}
   %\caption{Remember, \emph{never} use vertical lines in tables.}
   \label{tab:CV3}
\end{table}

\clearpage
\begin{table}[htbp]

   \centering
   \caption{Simulation under the Logit model: mean(sd) of AUC for effect identification. (a) Nonzero coefficients are equal to 0.5. (b) Nonzero coefficients are drawn from $\mathcal{U}[0.2,1]$. (c) Nonzero coefficients in dataset 2 are drawn from $\mathcal{U}[0.4,0.7]$, and those specific to datasets 1 and 3 are drawn from $\mathcal{U}[0.1,0.3]$ and $\mathcal{U}[0.8,1]$, respectively.} % requires the topcapt package
   \begin{tabular}{lcccccc} % Column formatting, @{} suppresses leading/trailing space
      \toprule
      %\multicolumn{2}{c}{correlation} \\
      & & $r=100$ & &  & $r=150$\\
      \cmidrule(r){2-4} \cmidrule(l){5-7}% Partial rule. (r) trims the line a little bit on the right; (l) & (lr) also possible
         & a &b & c &a &b& c \\
      \midrule
       \multicolumn{3}{l}{$(\rho_a,\rho_h,\rho_n)=(0.1,0,0.9)$} \\
        P.Lasso &0.561(0.02)&0.613(0.022)&0.552(0.018)&0.533(0.016)&0.582(0.017)&0.547(0.014)\\
       S.Lasso  &0.616(0.08)&0.618(0.079)&0.609(0.051)&0.541(0.084)&0.557(0.071)&0.538(0.074)\\
       CoFu &0.632(0.027)&0.646(0.026)&0.623(0.015)&0.579(0.011)&0.623(0.028)&0.594(0.021)\\
        \multicolumn{3}{l}{$(\rho_a,\rho_h,\rho_n)=(0.1,0.9,0)$} \\
        P.Lasso &0.624(0.012)&0.623(0.023)&0.619(0.038)&0.538(0.016)&0.569(0.021)&0.579(0.016)\\
       S.Lasso  &0.617(0.071)&0.587(0.08)&0.576(0.083)&0.521(0.076)&0.584(0.066)&0.581(0.067)\\
       CoFu &0.658(0.027)&0.656(0.03)&0.652(0.026)&0.606(0.015)&0.626(0.029)&0.636(0.067)\\
      \multicolumn{3}{l}{$(\rho_a,\rho_h,\rho_n)=(0.2,0.6,0.2)$} \\
       P.Lasso &0.621(0.029)&0.627(0.02)&0.622(0.024)&0.544(0.016)&0.592(0.023)&0.587(0.011)\\
       S.Lasso  &0.621(0.089)&0.584(0.089)&0.579(0.075)&0.56(0.095)&0.569(0.085)&0.553(0.098)\\
       CoFu &0.648(0.022)&0.65(0.024)&0.646(0.035)&0.61(0.028)&0.616(0.022)&0.627(0.023)\\
       \multicolumn{3}{l}{$(\rho_a,\rho_h,\rho_n)=(0.4,0.1,0.5)$} \\
       P.Lasso &0.645(0.03)&0.648(0.019)&0.631(0.031)&0.567(0.023)&0.595(0.024)&0.589(0.019)\\
       S.Lasso &0.595(0.086)&0.591(0.082)&0.614(0.067)&0.55(0.09)&0.544(0.088)&0.548(0.066)\\
       CoFu  &0.672(0.028)&0.668(0.014)&0.652(0.034)&0.619(0.025)&0.64(0.021)&0.639(0.019)\\
        \multicolumn{3}{l}{$(\rho_a,\rho_h,\rho_n)=(0.5,0.5,0)$} \\
      P.Lasso  &0.679(0.042)&0.687(0.022)&0.668(0.036)&0.614(0.028)&0.611(0.026)&0.623(0.017)\\
       S.Lasso   &0.625(0.081)&0.635(0.046)&0.62(0.072)&0.57(0.08)&0.586(0.084)&0.568(0.076)\\
       CoFu  &0.72(0.025)&0.699(0.028)&0.717(0.031)&0.668(0.028)&0.656(0.018)&0.654(0.032)\\
       \multicolumn{3}{l}{$(\rho_a,\rho_h,\rho_n)=(0.6,0.2,0.2)$} \\
        P.Lasso &0.672(0.022)&0.689(0.022)&0.69(0.015)&0.62(0.033)&0.647(0.027)&0.642(0.017)\\
       S.Lasso   &0.607(0.079)&0.611(0.083)&0.592(0.072)&0.533(0.07)&0.544(0.086)&0.543(0.08)\\
       CoFu  &0.678(0.021)&0.707(0.022)&0.707(0.017)&0.65(0.017)&0.676(0.026)&0.685(0.022)\\
         \multicolumn{3}{l}{$(\rho_a,\rho_h,\rho_n)=(0.9,0,0.1)$} \\
        P.Lasso  &0.77(0.016)&0.746(0.034)&0.774(0.022)&0.69(0.034)&0.695(0.03)&0.691(0.02)\\
       S.Lasso   &0.603(0.071)&0.613(0.071)&0.612(0.087)&0.563(0.087)&0.575(0.066)&0.558(0.072)\\
       CoFu  &0.764(0.028)&0.735(0.034)&0.778(0.033)&0.707(0.021)&0.709(0.015)&0.703(0.021)\\
      \bottomrule
   \end{tabular}
   \label{tab:log1}
\end{table}

\clearpage
\begin{table}[htbp]

   \centering
   \caption{Simulation under the Logit model: mean(sd) of AUC for community differentiation.  (a) Nonzero coefficients are equal to 0.5. (b) Nonzero coefficients are drawn from $\mathcal{U}[0.2,1]$. (c) Nonzero coefficients in dataset 2 are drawn from $\mathcal{U}[0.4,0.7]$, and those specific to datasets 1 and 3 are drawn from $\mathcal{U}[0.1,0.3]$ and $\mathcal{U}[0.8,1]$, respectively.} % requires the topcapt package
   \begin{tabular}{lcccccc} % Column formatting, @{} suppresses leading/trailing space
      \toprule
      %\multicolumn{2}{c}{correlation} \\
      & & $r=100$ & &  & $r=150$\\
      \cmidrule(r){2-4} \cmidrule(l){5-7}% Partial rule. (r) trims the line a little bit on the right; (l) & (lr) also possible
         & a & b & c &a &b& c \\
      \midrule
        \multicolumn{3}{l}{$(\rho_a,\rho_h,\rho_n)=(0.1,0,0.9)$} \\
       S.Lasso  &0.591(0.077)&0.613(0.083)&0.537(0.081)&0.585(0.071)&0.606(0.108)&0.545(0.071)\\
       CoFu &0.663(0.067)&0.659(0.075)&0.623(0.065)&0.631(0.067)&0.687(0.063)&0.639(0.058)\\
       \multicolumn{3}{l}{$(\rho_a,\rho_h,\rho_n)=(0.1,0.9,0.1)$} \\
       S.Lasso  &0.569(0.064)&0.605(0.09)&0.604(0.073)&0.536(0.097)&0.576(0.081)&0.539(0.138)\\
       CoFu &0.639(0.052)&0.688(0.053)&0.661(0.056)&0.59(0.071)&0.623(0.07)&0.596(0.083)\\
      \multicolumn{3}{l}{$(\rho_a,\rho_h,\rho_n)=(0.2,0.6,0.2)$} \\
       S.Lasso  &0.547(0.075)&0.577(0.08)&0.586(0.084)&0.52(0.078)&0.573(0.108)&0.557(0.06)\\
       CoFu   &0.627(0.073)&0.633(0.078)&0.645(0.079)&0.608(0.08)&0.644(0.062)&0.615(0.068)\\
       \multicolumn{3}{l}{$(\rho_a,\rho_h,\rho_n)=(0.4,0.1,0.5)$} \\
       S.Lasso  &0.536(0.08)&0.594(0.052)&0.523(0.134)&0.529(0.177)&0.544(0.056)&0.551(0.085)\\
       CoFu &0.627(0.073)&0.648(0.049)&0.619(0.102)&0.612(0.112)&0.626(0.064)&0.628(0.076)\\
       \multicolumn{3}{l}{$(\rho_a,\rho_h,\rho_n)=(0.5,0.5,0)$} \\
       S.Lasso  &0.516(0.076)&0.513(0.079)&0.512(0.063)&0.568(0.108)&0.556(0.054)&0.541(0.051)\\
       CoFu  &0.595(0.048)&0.614(0.048)&0.633(0.069)&0.602(0.086)&0.6(0.064)&0.619(0.062)\\
       \multicolumn{3}{l}{$(\rho_a,\rho_h,\rho_n)=(0.6,0.2,0.2)$} \\
       S.Lasso  &0.522(0.076)&0.558(0.096)&0.498(0.06)&0.537(0.081)&0.564(0.085)&0.452(0.081)\\
       CoFu &0.652(0.043)&0.647(0.057)&0.685(0.06)&0.635(0.062)&0.644(0.069)&0.636(0.051)\\
        \multicolumn{3}{l}{$(\rho_a,\rho_h,\rho_n)=(0.9,0,0.1)$} \\
       S.Lasso  &0.584(0.114)&0.596(0.088)&0.538(0.146)&0.598(0.168)&0.577(0.107)&0.565(0.107)\\
       CoFu &0.71(0.087)&0.707(0.042)&0.699(0.065)&0.704(0.121)&0.688(0.056)&0.676(0.064)\\
      \bottomrule
   \end{tabular}
   %\caption{Remember, \emph{never} use vertical lines in tables.}
   \label{tab:log2}
\end{table}
%\end{sidewaystable}

\begin{figure}[h] %
\centering
   \includegraphics[width=6in]{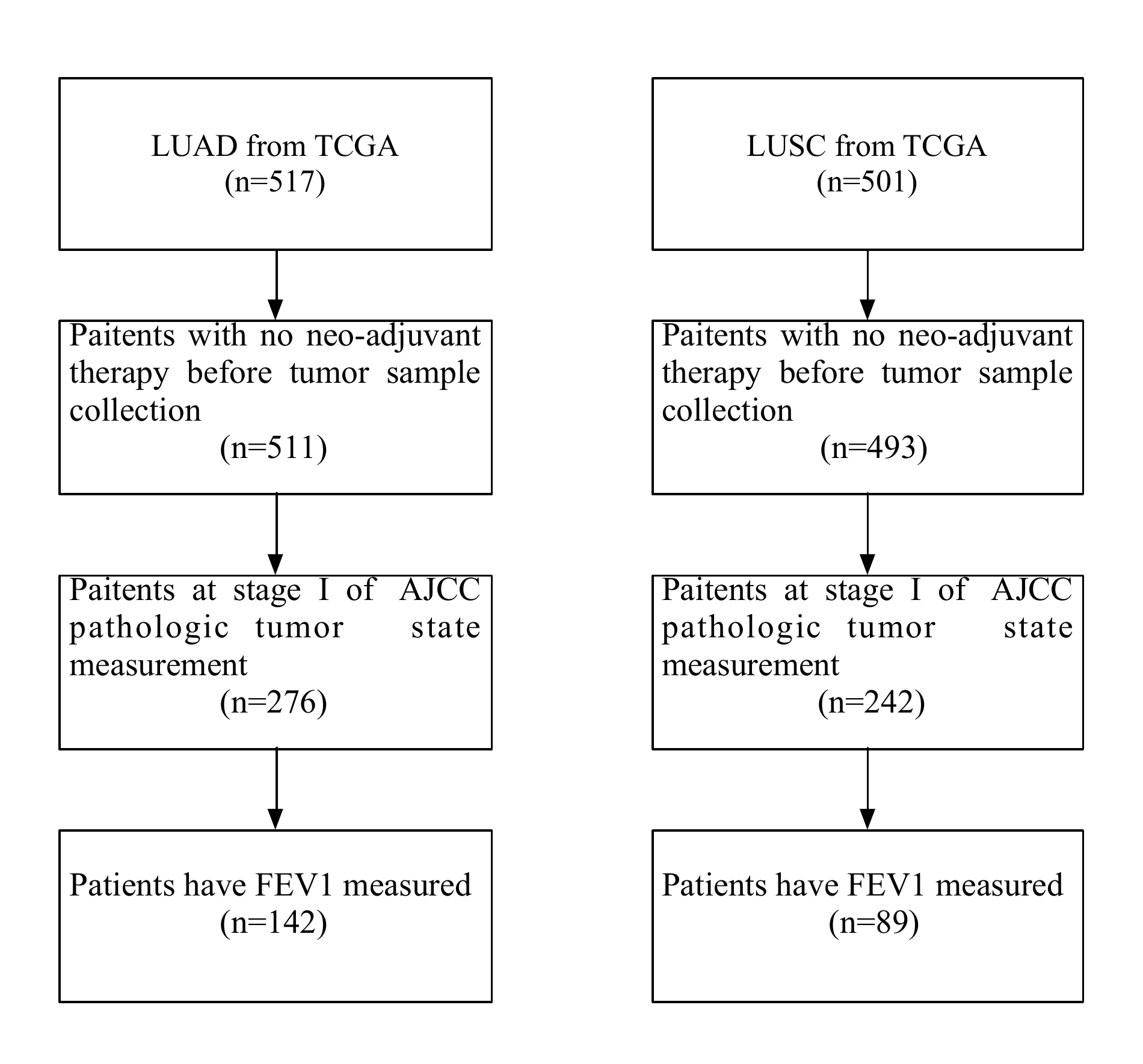}
   \caption{Flowcharts of sample selection for the analysis of lung cancer data (left: LUAD, right: LUSC).\label{fig:flowchart}}
\end{figure}

\clearpage
\begin{table}[htbp]%\scriptsize
   \centering
   \caption{Simulation based on the SKCM data: mean(sd) of AUC.} % requires the topcapt package
   \begin{tabular}{cccccc} % Column formatting, @{} suppresses leading/trailing space
      \toprule
        &community& &  & individual effect\\
      \cmidrule(r){2-3} \cmidrule(l){4-6}% Partial rule. (r) trims the line a little bit on the right; (l) & (lr) also possible
       &S.Lasso&CoFu &P.Lasso&S.Lasso &CoFu\\
      \midrule
     \multicolumn{3}{l}{$(\rho_a,\rho_h,\rho_n)=(0.1,0,0.9)$} \\
       &0.772(0.051)&0.882(0.032)&0.576(0.011)&0.659(0.012) & 0.673(0.01)\\
      \multicolumn{3}{l}{$(\rho_a,\rho_h,\rho_n)=(0.1,0.9,0)$} \\
       &0.758(0.081)&0.819(0.035)&0.625(0.009)&0.651(0.016) & 0.706(0.012)\\
      \multicolumn{3}{l}{$(\rho_a,\rho_h,\rho_n)=(0.2,0.6,0.2)$} \\
      &0.514(0.062) & 0.631(0.032)& 0.656(0.012) & 0.655(0.009) & 0.679(0.008)\\
       \multicolumn{3}{l}{$(\rho_a,\rho_h,\rho_n)=(0.4,0.1,0.5)$} \\
       &0.6(0.048) & 0.71(0.042)&0.686(0.018) & 0.661(0.014) & 0.723(0.011)\\
       \multicolumn{3}{l}{$(\rho_a,\rho_h,\rho_n)=(0.5,0.5,0)$} \\
       &0.563(0.055) & 0.676(0.038)&0.665(0.009) & 0.658(0.011) &0.724(0.006) \\
       \multicolumn{3}{l}{$(\rho_a,\rho_h,\rho_n)=(0.6,0.2,0.2)$} \\
      &0.633(0.046) & 0.712(0.043)&0.706(0.013) & 0.659(0.008) &0.704(0.014)\\
        \multicolumn{3}{l}{$(\rho_a,\rho_h,\rho_n)=(0.9,0,0.1)$} \\
     &0.566(0.066)&0.829(0.046)&0.752(0.015)&0.66(0.012) & 0.749(0.01)\\
      \bottomrule
   \end{tabular}
   %\caption{Remember, \emph{never} use vertical lines in tables.}
   \label{tab:realdata}
\end{table}

\clearpage
\begin{figure}[h] %
   \includegraphics[width=7.5in]{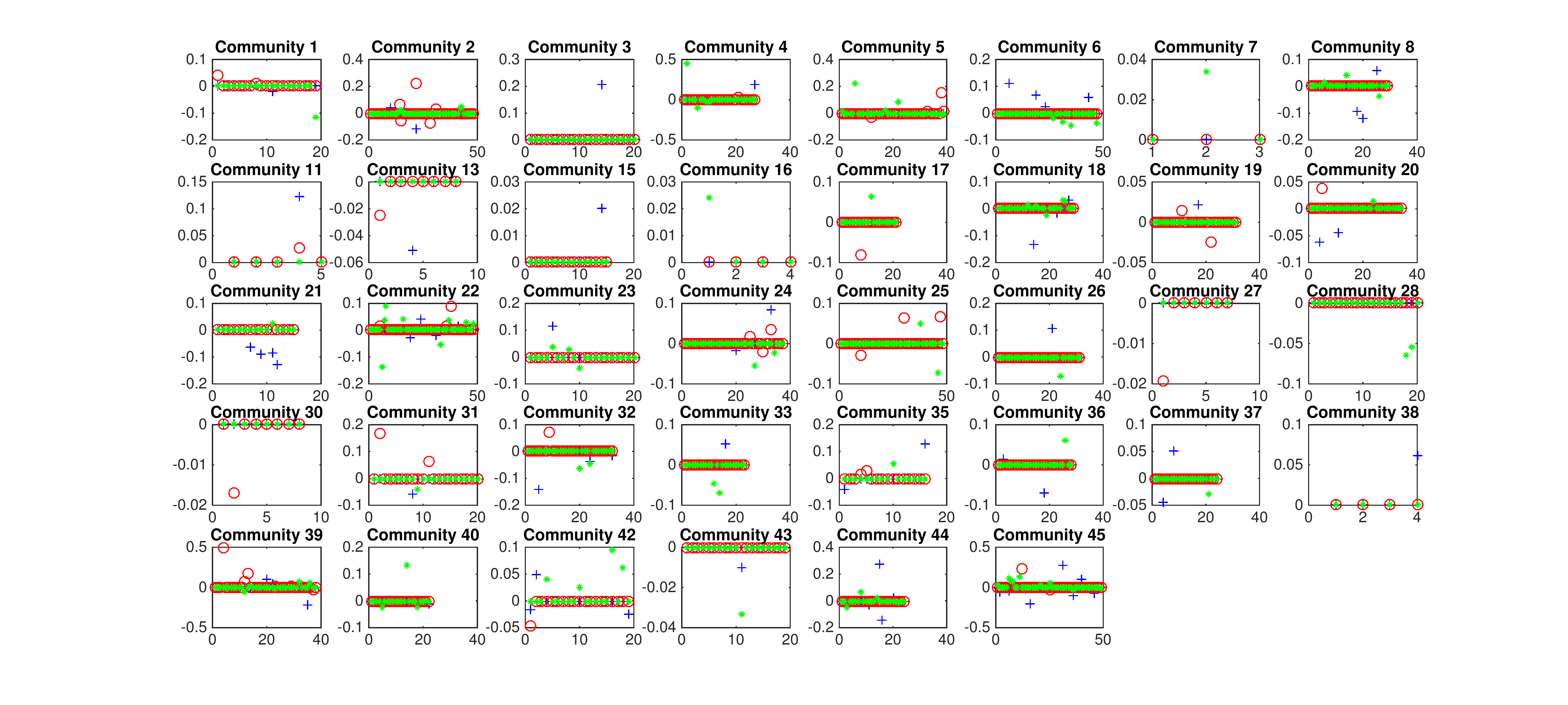}
   \caption{Analysis of the TCGA SKCM data using S.Lasso. Blue crosses correspond to stage I, red circles to stage II, and green filled circles to stages III and IV.\color{black}}
   \label{fig:lasso_Mela}
\end{figure}

\clearpage
\begin{figure}[h] %
   \includegraphics[width=7.5in]{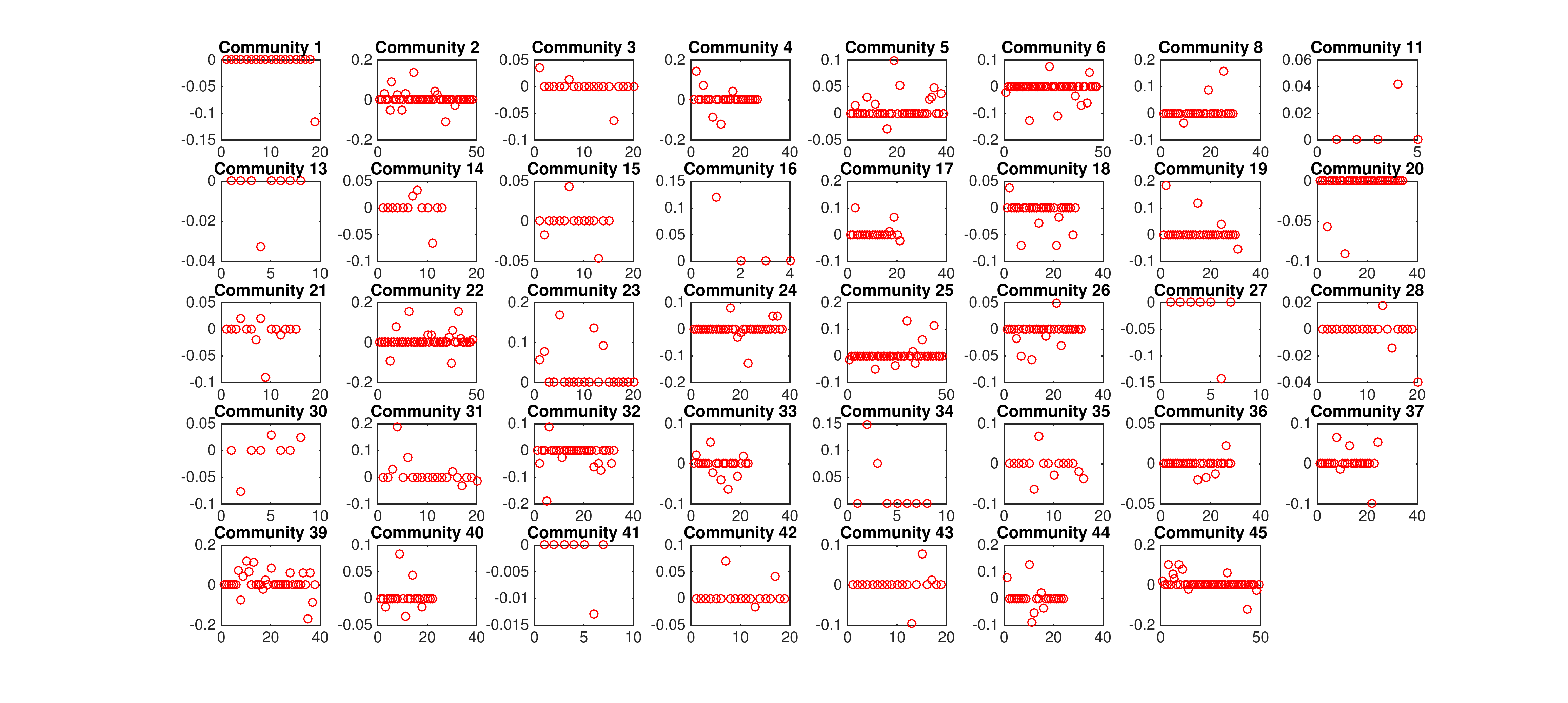}
   \caption{Analysis of the TCGA SKCM data using P.Lasso. }
   \label{fig:single_Mela}
\end{figure}

\clearpage
\begin{figure}[h] %
   \includegraphics[width=7.5in]{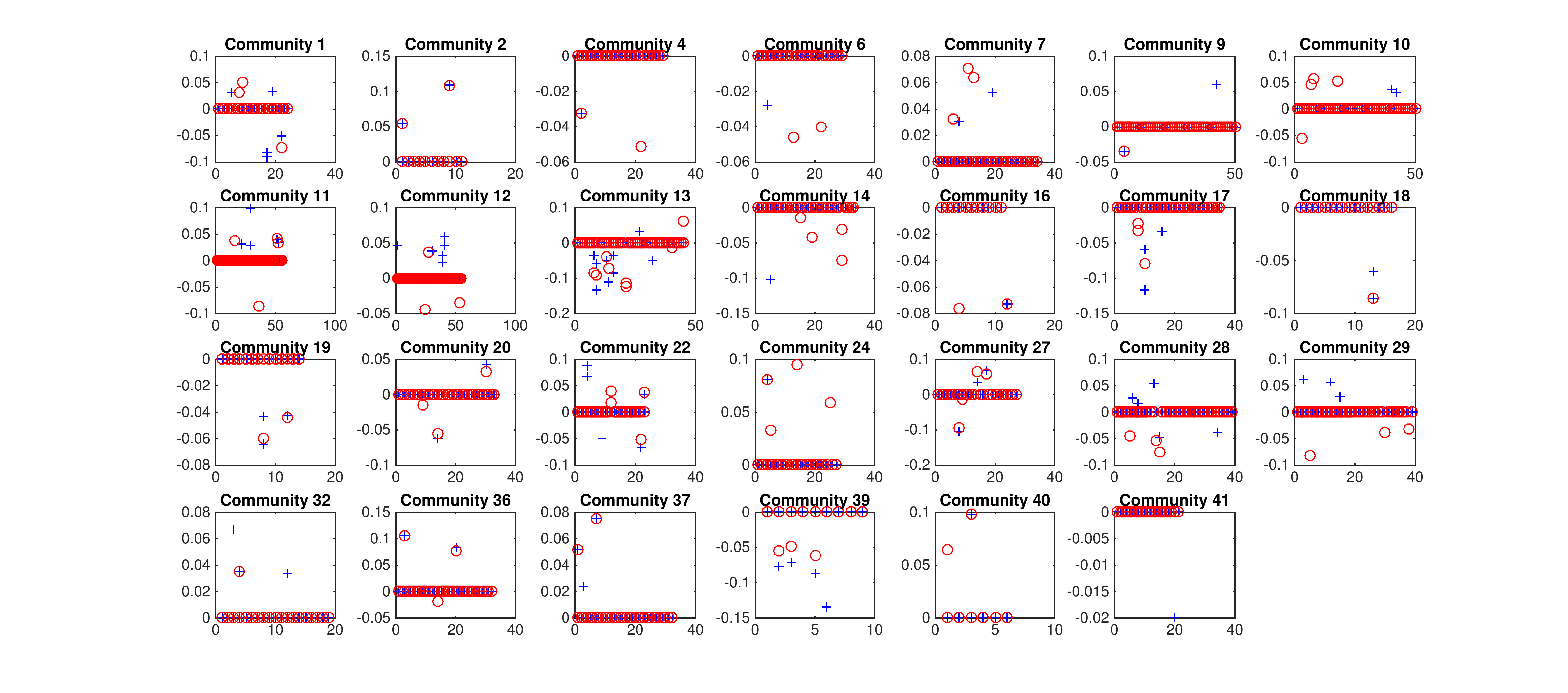}
   \caption{Analysis of the TCGA lung cancer data using S.Lasso. Blue crosses correspond to LUAD 
   and red circles to LUSC. }
   \label{fig:lasso_Lung}
\end{figure}

\clearpage
\begin{figure}[h] %
   \includegraphics[width=7.5in]{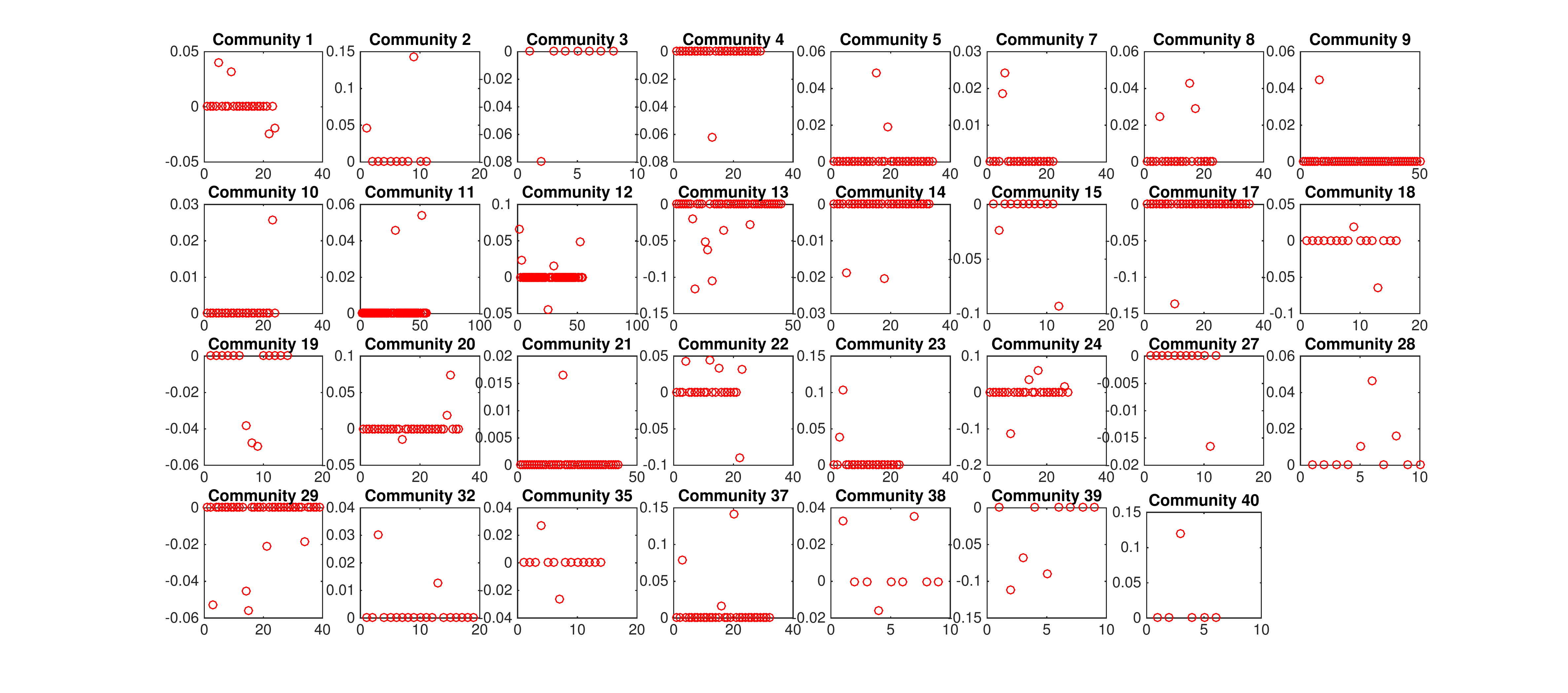}
   \caption{Analysis of the TCGA lung cancer data using P.Lasso. }
   \label{fig:single_Lung}
\end{figure}

\clearpage
\begin{figure}[h] %
   \centering
   \includegraphics[width=7in]{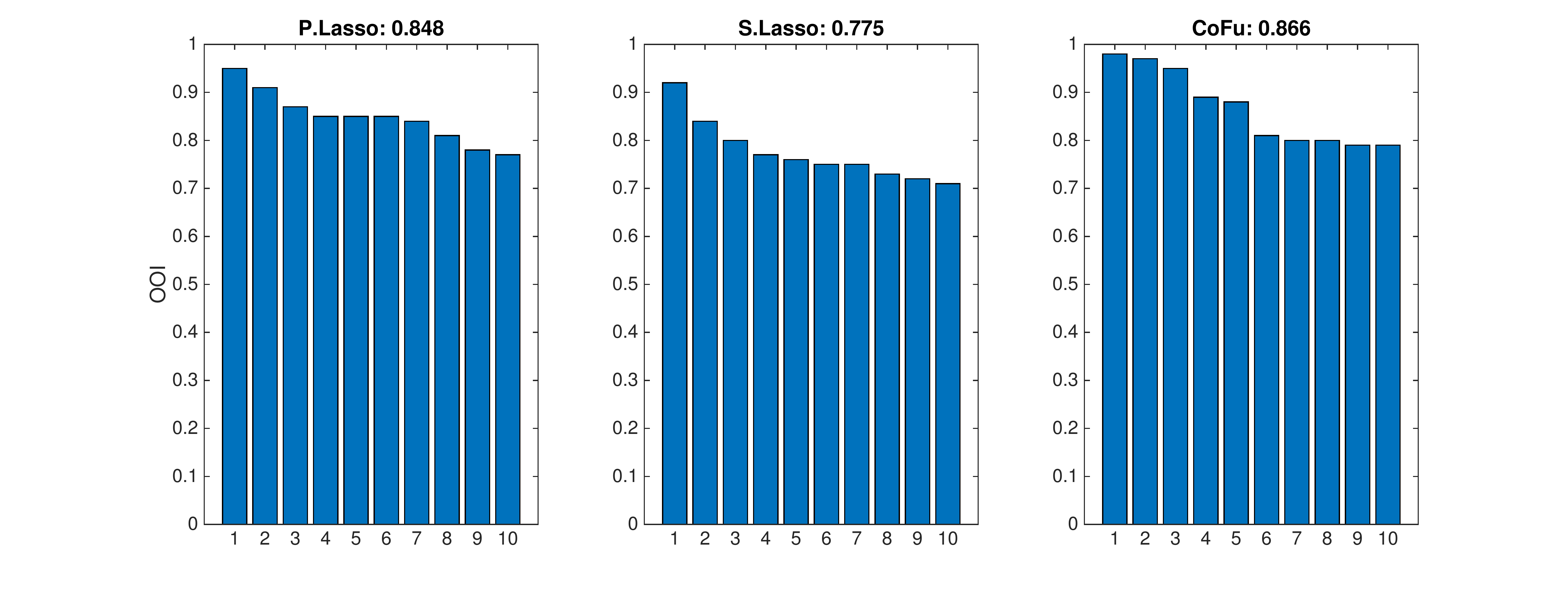}\\
   \includegraphics[width=7in]{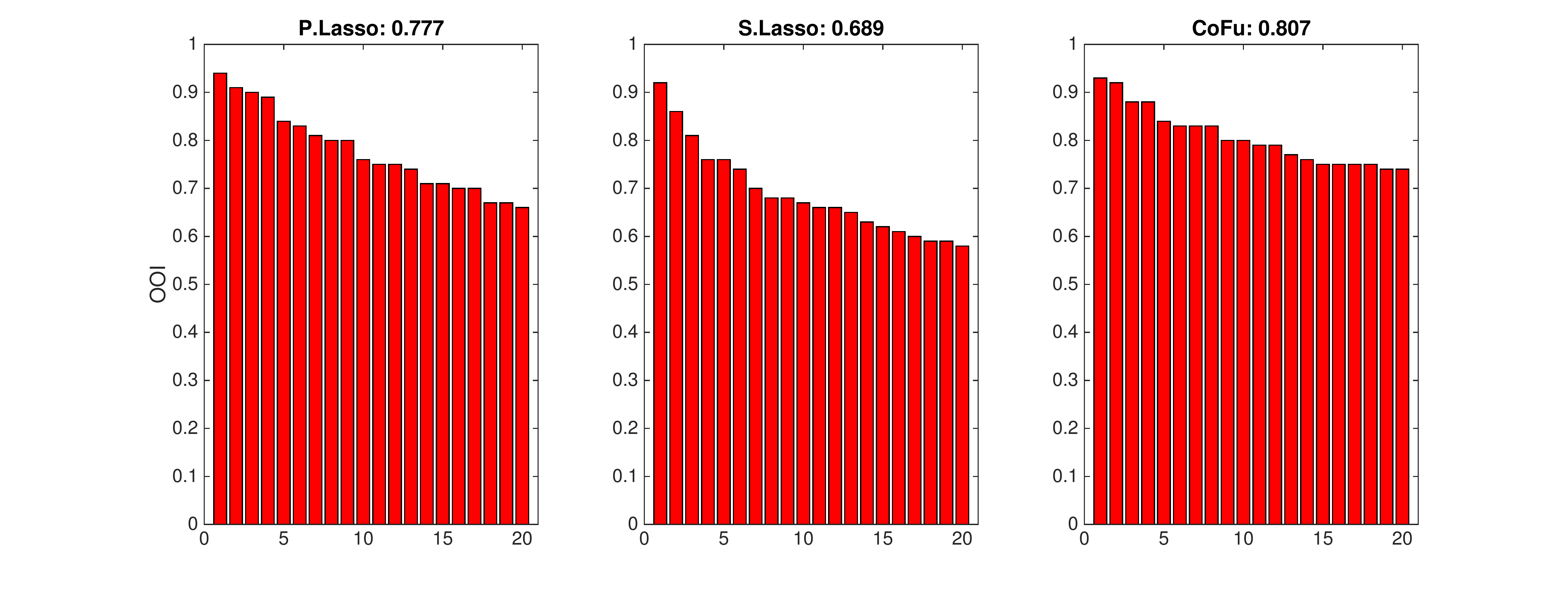}\\
   \caption{OOI in data analysis. Top: SKCM. Bottom: lung cancer data.}
   \label{fig:OOI}
\end{figure}

\section{Computation of AUC in simulation}
\label{sec:AUC}
For the proposed CoFu method, we fit the model under $20\times6$ $[\lambda_1, \lambda_2]$ values, where $\lambda_2 \in (0,0.001,0.01,0.1,1,10)$. Let $\lambda_1^{\max}$ be the minimal $\lambda_1$ such that all regression coefficients shrink to zero. We generate a sequence of $\lambda_1$ with length 20 that are equally-spaced in logarithm from $0.001\lambda_1^{\max}$ to $\lambda_1^{\max}$. We calculate the TPR and FPR for each $[\lambda_1, \lambda_2]$, and extract the envelope, which is the ROC curve. From this, we obtain the AUC. For the two alternatives, the calculation of the ROC and AUC can be conducted in a similar manner.

\section{Examination of censoring in the analysis of lung cancer data}
\label{sec:censoring}
We take a closer look at the vital status in the analysis of the lung cancer data. The contingency table is shown below. For the LUAD data, stage I has a higher percentage of censoring, whereas for the LUSC data, there is no significant association between vital status and stage. In our analysis, we focus on the continuous FEV1 variable, which is not subject to censoring. All subjects with FEV1 measurements are used in analysis, regardless of their vital status. Censoring is expected to play a more direct role in the analysis of prognosis.

\clearpage
\begin{table}[htbp]
   \centering
   \caption{Contingency table for LUAD and LUSC} % requires the topcapt package
   \begin{tabular}{lcccccc} % Column formatting, @{} suppresses leading/trailing space
      \toprule
      %\multicolumn{2}{c}{correlation} \\
      & &LUAD & & & LUSC & \\
      \cmidrule(r){2-4} \cmidrule(r){5-7} % Partial rule. (r) trims the line a little bit on the right; (l) & (lr) also possible
         & Stage I &Stage II & Stage III   & Stage I &Stage II & Stage III\\
      \midrule
       %\multicolumn{3}{l}{$(\rho_a,\rho_h,\rho_n)=(0.1,0,0.9)$} \\
        Deceased & 68 & 53 & 61 & 99 & 63 & 48\\
        Alive & 208 & 68 & 45 & 143 & 94 & 42\\
      \bottomrule
   \end{tabular}
   \label{tab:con1}
\end{table}

\color{black}

\end{document}